\documentclass[a4paper,11pt]{article}

\usepackage[boxruled]{algorithm2e}
\SetKw{Continue}{continue}
\SetKw{Break}{break}
\SetKw{True}{true}

\usepackage{amsmath}
\usepackage{amsfonts}

\usepackage{amssymb}
\usepackage{tikz}
\usepackage{mathrsfs}
\usepackage[colorinlistoftodos]{todonotes}

\usepackage{xcolor}
\usepackage{hyperref}
\hypersetup{
    colorlinks,
    linkcolor={green!50!black},
    citecolor={blue!50!black},
    urlcolor={blue!80!black}
}

\usepackage[inline]{enumitem}

\usepackage{amsthm}
\usepackage{cases}
\usepackage{stmaryrd}

\usepackage[capitalize]{cleveref}

\newtheorem{theorem}{Theorem}
\newtheorem{lemma}[theorem]{Lemma}
\newtheorem{corollary}[theorem]{Corollary}
\newtheorem{definition}[theorem]{Definition}

\newtheorem{claim}[theorem]{Claim}

\newtheoremstyle{restate}{}{}{\itshape}{}{\bfseries}{~(restated).}{.5em}{\thmnote{#3}}
\theoremstyle{restate}

\DeclareMathOperator*{\argmax}{arg\,max}
\DeclareMathOperator*{\argmin}{arg\,min}

\newcommand{\cD}{\mathcal{D}}

\newcommand{\cA}{\mathcal{A}}
\newcommand{\cB}{\mathcal{B}}

\DeclareMathOperator{\Unif}{Unif}
\DeclareMathOperator*{\E}{\mathbb{E}}
\DeclareMathOperator*{\Var}{Var}
\newcommand{\eps}{\epsilon}
\newcommand{\R}{\mathbb{R}}
\newcommand{\Norm}[1]{\left\| #1 \right\|}
\newcommand{\Abs}[1]{\left| #1 \right|}
\newcommand{\inner}[2]{\left\langle #1, #2 \right\rangle}
\newcommand{\match}{\leftrightsquigarrow}

\usepackage{stackengine}

\newcommand{\ee}{\stackon[2pt]{\^{e}}{\~{}}}

\DeclareMathOperator{\poly}{poly}
\DeclareMathOperator{\polylog}{polylog}
\DeclareMathOperator{\nnz}{nnz}

\newcommand{\setzero}[1]{\{0,\dots,#1\}}

\newcommand{\modify}[1]{{#1}}
\newcommand{\smallmodify}[1]{{#1}}
\usepackage[margin=1in]{geometry}

\title{Sparsity-Dimension Trade-Offs \\for Oblivious Subspace Embeddings} 

\author{Yi Li\thanks{Supported in part by Singapore Ministry of Education (AcRF) Tier 1 grant RG75/21 and Tier 2 grant MOE-T2EP20122-0001.}
\\
School of Physical and Mathematical Sciences\\
Nanyang Technological University\\
\texttt{yili@ntu.edu.sg}
\and 
Mingmou Liu\thanks{%
Supported by the VILLUM Foundation grant 16582. Part of this work was done when the author was a research fellow at the Nanyang Technological University, supported by Singapore Ministry of Education (AcRF) Tier 1 grant RG75/21.}\\
Basic Algorithms Research Copenhagen\\
University of Copenhagen \\
\texttt{mili@di.ku.dk}
}

\date{}

\begin{document}

\maketitle

\begin{abstract}
An oblivious subspace embedding (OSE), characterized by parameters $m,n,d,\epsilon,\delta$, is a random matrix $\Pi\in \mathbb{R}^{m\times n}$ such that for any $d$-dimensional subspace $T\subseteq \mathbb{R}^n$, $\Pr_\Pi[\forall x\in T, (1-\epsilon)\|x\|_2 \leq \|\Pi x\|_2\leq (1+\epsilon)\|x\|_2] \geq 1-\delta$.
When an OSE has $s\le 1/2.001\epsilon$ nonzero entries in each column, we show it must hold that $m = \Omega\left(d^2/( \epsilon^2s^{1+O(\delta)})\right)$, which is the first lower bound with multiplicative factors of $d^2$ and $1/\epsilon$, improving on the previous $\Omega\left(d^2/s^{O(\delta)}\right)$ lower bound due to Li and Liu (PODS 2022). 
When an OSE has $s=\Omega(\log(1/\epsilon)/\epsilon)$ nonzero entries in each column, we show it must hold that $m = \Omega\left((d/\epsilon)^{1+1/4.001\epsilon s}/s^{O(\delta)}\right)$, which is the first lower bound with multiplicative factors of $d$ and $1/\epsilon$, improving on the previous $\Omega\left(d^{1+1/(16\epsilon s+4)}\right)$ lower bound due to Nelson and Nguy\ee{}n (ICALP 2014). This second result is a special case of a more general trade-off among $d,\epsilon,s,\delta$ and $m$. 
\end{abstract}

\section{Introduction}

Subspace embedding is an essential dimensionality reduction tool for processing large datasets, such as clustering~\cite{BZMD,Cohen15}, performing correlation analysis~\cite{ABTZ}, and solving basic linear algebraic problems like regression and low-rank approximation~\cite{CW17}. 
Formally, given a subspace $T\subseteq \R^n$, a linear map $\Pi:\R^n\to \R^m$ (or, equivalently, a matrix $\Pi\in\R^{m\times n}$) is called an $\eps$-subspace-embedding for $T$ if $(1-\epsilon)\Norm{x}_2 \leq \Norm{\Pi x}_2 \leq (1+\epsilon)\Norm{x}_2$ for all $x\in T$ simultaneously. In this definition, $\Pi$ can be dependent on the subspace $T$; however, in many applications, the data can be dynamically updated (and thus $T$ changes), so it is preferable to have a $\Pi$ that is independent of $T$, in which case $\Pi$ is called an oblivious subspace embedding. It is clear that a fixed $\Pi$ cannot be a subspace embedding for all subspaces $T$; a more reasonable definition thus includes randomness in $\Pi$. A formal definition is given below.

\begin{definition}
An $(n,d,\eps,\delta)$-oblivious-subspace-embedding is a random matrix $\Pi\in \R^{m\times n}$ such that for any fixed $d$-dimensional subspace $T\subseteq \R^n$, it holds that
\begin{equation}\label{eqn:OSE}
\Pr\left[\forall x\in T, (1-\eps)\|x\|_2 \leq \|\Pi x\|_2 \leq (1+\eps)\|x\|_2\right] \geq 1 - \delta.
\end{equation}
\end{definition}

In most applications, the subspace $T$ is the column space of a data matrix $A\in\R^{n\times d}$,  so \cref{eqn:OSE} can also be written as 
\begin{equation}\label{eqn:SE_matrix}
\Pr \left[\forall x\in \R^d, (1-\eps)\|Ax\|_2 \leq \|\Pi A x\|_2 \leq (1+\eps)\|Ax\|_2\right] \geq 1 - \delta.
\end{equation}

When $n$ and $d$ are clear from the context, we may omit them in the parameters. We  shall consider only oblivious subspace embeddings, so we will also drop the word ``oblivious'' and simply say ``$(\eps,\delta)$-subspace-embedding''.

A typical application of oblivious subspace embedding is linear regression, i.e., solving $\min_{x\in\R^d} \Norm{Ax-b}_2$ for given $A\in \mathbb{R}^{n\times d}$ and $b\in \R^n$ with $n\gg d$. 
In products, linear regression solvers have been included as native functions in some database management systems (DBMS) and various machine learning packages for databases (e.g. MADlib~\cite{10.14778/2367502.2367510}, MLlib~\cite{mllib}, Oracle Database~\cite{oracle}, and SystemML~\cite{systemml}). In research, linear regression appears as a vehicle for optimizing designs when studying joined tables~\cite{JSWY21,10.5555/1124191.1124299,10.1145/2882903.2882939} and privacy preservation~\cite{10.1145/3034786.3034795}.
To see how a subspace embedding plays its role, suppose that $\Pi$ is an $(\eps,\delta)$-subspace-embedding for the column space of the concatenated matrix $(A \ b)$. This subspace has dimension at most $d+1$, so $\Pi$ can have only $\poly(d/\eps)\ll n$ rows. The subspace embedding property of $\Pi$ implies that $\Norm{\Pi Ax-\Pi b}_2 = \Norm{\Pi(Ax-b)}_2 =  (1\pm\eps)\Norm{Ax-b}_2$ for all $x$. As a result, the original large-scale problem $\Norm{Ax-b}_2$ is reduced to a much smaller one $\Norm{\Pi Ax-\Pi b}_2$, which can be solved much faster.

The goal of designing $\Pi$ is to keep the number $m$ of rows to a minimum and to make applying $\Pi$ to $A$, i.e.\ computing the matrix product $\Pi A$, as computationally efficient as possible.  The smallest attainable $m$ is $m = \Theta((d + \log(1/\delta))/\eps^2)$~\cite{NN14}, but such constructions are dense matrices, leading to a long time to compute $\Pi A$. There have been two routes to improve the runtime. 
The first is to seek a more structured $\Pi$ that allows for faster computation, such as involving the discrete Fourier transform so that the Fast Fourier Transform can be applied; 
the other is to seek a sparser $\Pi$ with a small number $s$ of nonzero entries in each column so that computing $\Pi A$ takes only $O(s\cdot\nnz(A))$ time, where $\nnz(A)$ denotes the number of nonzero entries of $A$. We consider the second route, i.e., sparse oblivious subspace embeddings, in this paper.

The sparsest case, in which $s=1$, is now completely understood. The classical Count-Sketch construction has exactly one nonzero entry in each column. Its value is chosen uniformly at random from $\{-1,1\}$ and its position uniformly at random in that column. 
It is known that $m = O(d^2/(\eps^2\delta))$ rows are sufficient to provide an $(\eps,\delta)$-subspace-embedding~\cite{NN13:OSNAP}. The quadratic dependence on $d$ has been known to be tight for several years~\cite{NN13:LB}, but nothing was known when combined with the dependence on $\eps$ as a multiplicative factor until recently Li and Liu showed a fully tight bound of $m = \Omega(d^2/(\eps^2\delta))$~\cite{LL22}.

The current understanding of larger $s$ is much less satisfying. A classical construction with larger $s$ is OSNAP~\cite{NN13:OSNAP,Cohen16}, which exhibits (i) $m=O(d \log(d/\delta) /\eps^2)$ when $s=\Theta(\log(d/\delta)/\eps)$, or (ii) $m=O(d^{1+\gamma}\log(d/\delta)/\eps^2)$ when $s=\Theta(1/(\gamma\eps))$ for any constant $\gamma>0$. 
Recall the general lower bound $m = \Omega((d + \log(1/\delta))/\eps^2)$~\cite{NN14}, we see that $m$ in case (i) is tight up to a logarithmic factor. In case (ii), as $s$ decreases by a logarithmic factor, $m$ significantly increases while still maintaining a subquadratic dependence on $d$, namely, $d^{1+1/\Theta(\eps s)}$. This form matches the lower bound of $\Omega(d^{1+1/(16\eps s+4)})$~\cite{NN14}\footnote{The constants such as $16$ and $4$ are not specified in the theorem statement in~\cite{NN14} but can be determined through a careful examination of the proof.}.
On the other hand, when $s$ is a constant factor smaller, say, $s \leq \alpha/\eps$ for some constant $\alpha > 0$, a quadratic dependence on $d$ becomes necessary and it is known that $m = \Omega(\eps^2 d^2)$ for constant $\delta$~\cite{NN14}. The dependence on $\eps$ was later improved to $m = \Omega(\eps^{O(\delta)}d^2)$~\cite{LL22}. Recall that Count-Sketch ($s=1$) attains the tight bound $m = \Theta(d^2/\eps^2)$, no construction with a substantially better $m$ is known for $s\leq \alpha/\eps$, leaving a large gap between the upper and lower bounds in this regime.

We conjecture that a tight lower bound $m = \Omega(d^2/\eps^2)$ exists when $s\leq \alpha/\eps$ and a lower bound $m=\Omega(d^{1+1/\Theta(\eps s)}/\eps^2)$ exists when $s\ge \alpha/\eps$, so there is a sharp transition for $s\in(\eps^{-1},O(\eps^{-1}\log d))$. As the first step towards this goal, we ask the following questions:
\begin{center}
\noindent\textit{
Is it possible to obtain a lower bound including both $d^2$ (resp. $d^{1+1/\Theta(\eps s)}$) and $1/\eps$ as multiplicative factors, such as $\Omega(d^2/\eps^{0.1})$ (resp. $\Omega(d^{1+1/\Theta(\eps s)}/\eps^{0.1})$), when $s\leq \alpha/\eps$ (resp. $s\geq \alpha/\eps$)?
}
\end{center}

\paragraph{Our Results.}
In this paper, we answer the questions above affirmatively by providing two trade-offs between $m$ and $s$. These are the first lower bounds with multiplicative factors of both $d^{1+1/\Theta(\eps s)}$ and $1/\eps$.

The first result states that any $(\eps,\delta)$-subspace-embedding with column sparsity at most $(1/2-o(1))\eps$ must have $\Omega(d^2/(s^{1+O(\delta)}\eps^{2}\polylog s))$ rows. The result is formally stated below.

\begin{theorem}\label{thm:main}
There exist absolute constants $\eps_0, \delta_0, c_0, K_0, K_1 > 0$ such that the following holds.
For all $\eps \in (0,\eps_0)$, $\delta \in (0,\delta_0)$, $d\ge 1/\eps^2$ and $n\ge K_0 d^{2}/(\eps^2\delta)$, 
any  $(\eps,\delta)$-subspace-embedding $\Pi$ must have at least $$m\ge c_0 d^2/(\eps^{2}s^{1+K_1\delta}\log^{13} s)$$ rows
if
the column sparsity $s$ of $\Pi$ satisfies $3\leq s \leq \frac{(1-\eps)^2+(2\eps-\eps^2)/\log s+1/\log^2s }{2\eps}$.
\end{theorem}
Note that the upper bound for $s$ is $(1/2-o(1))\eps$ if $\eps=o(1)$ and $s=\omega(1)$. %

\bigskip

Our second result is a general lower bound for sparse OSE: 
any OSE with column sparsity $s$ must have $\Omega((d/s\eps)^{1+1/(\lfloor (4+o(1))\eps s\rfloor+1)}s^{1-O(\delta)}/\polylog s)$ rows.
 The result is formally stated below.
\begin{theorem}\label{thm:main2}
There exist absolute constants $\eps_0, \delta_0, c_0, K_0, K_1, K_2> 0$ such that the following holds.
For all $\eps \in (0,\eps_0)$, $\delta \in (0,\delta_0)$, $d\ge 1/\eps^7$ and $n\ge K_0 d^{2}/(\eps^2\delta)$, 
any  $(\eps,\delta)$-subspace-embedding $\Pi$ must have at least 
    \[
        m \geq \frac{c_0\cdot\theta^{1/(\chi-1)}}{s^{K_1\delta\chi/(\chi -1)}\log^{6/{(\chi-1)}+2}s}\cdot\left(\frac{d}{\eps}\right)^{1+1/(\chi-1)},
    \]
    rows, 
if the column sparsity $s$ of $\Pi$ satisfies $3\leq s \leq K_2{\eps}^{-1}{\log d}$,
    where $\chi = \lfloor (4\eps+\theta)/\theta(1-1/\log s)\rfloor+1$ and $\theta\ge (1-\eps)^2/s-1/s\log s$.
\end{theorem}
Note that $4\eps/\theta= 4\eps s (1+2\eps/(1-\eps)^2+O(1/\log s))$, and $\chi=\lfloor (4+o(1))\eps s\rfloor+2$ if $s=\omega(1)$ and $\eps=o(1)$.
Thus for $\eps\le 1/202$ and $s=\Omega(\log(1/\eps)/\eps)\cap O((\log d)/\eps)$, our lower bound is at least
\[
m=\Omega\left(\left(\frac{d}{\eps}\right)^{1+\frac{1}{4(1+2\eps)\eps s}}\cdot{s^{-O(\delta)}}\right),
\]
which is better than the previous $\Omega\left(d^{1+1/(16\eps s+4)}\right)$ lower bound of Nelson and Nguy\stackon[2pt]{\^{e}}{\~{}}n~\cite{NN14} in the dependence of both $d$ and $\eps$.

We remark that the first lower bound is significantly stronger than the second one when $1/4\eps<s<1/2\eps$.
This is because the lower bounds are derived from contradictions through different principles.
We further remark that the same lower bound can be obtained for $s=1,2$ (except that $\log s$ is replaced with some constant) with a minor modification to our proof.

\section{Technique overview}
We begin by applying Yao's minimax principle to construct a matrix distribution $\cD$ (which is called the \emph{hard instance}) and to show that any deterministic $\Pi$ that is an $(\eps,\delta)$-subspace-embedding for the column space of a random matrix $U\sim\cD$, i.e. \cref{eqn:SE_matrix} holds for $U\sim\cD$ and deterministic $\Pi$, must have many rows.

We adopt the hard instance $\widetilde{\cD}$ proposed in \cite{LL22}, which is formally given in \cref{def:D_beta}. 
Informally speaking, $U\sim\widetilde{\cD}$ has a random parameter $\beta\in(0,1]$ and $U=VW$.
When the random parameter $\beta$ is fixed to some specified value, we denote $U\sim\cD_\beta$. %
Intuitively, the setting forces $\Pi$ to be an OSE for $U\sim\cD_\beta$ with most $\beta\in(0,1]$.
The matrix $V$ has $d/\beta$ columns and each column $(\Pi V)_{\star,i}$ is a copy of a random column of $\Pi$.\footnote{For a matrix $A$, we denote its $i$-th column vector by $A_{\star, i}$.}
The matrix $W$ is a random matrix which guarantees that if two column vectors of $\Pi V$ have a large inner product in absolute value, then $\Pi$ is unlikely to be a subspace embedding for $U=VW$ (see \cref{lemma: inner product implies anticoncentration}). Therefore, our proof for the sparse case (so do the proofs in \cite{NN14,LL22}) is all about finding a large inner product among the columns of $\Pi V$.

Specifically, the proof for \cref{thm:main} is devoted to finding a pair of columns of $\Pi V$ whose inner product is more than $2\eps$. The random $W$ randomizes $\|\Pi Uv\|_2^2$ for some unit vector $v$, causing $\|\Pi Uv\|_2^2$ to fluctuate over a range of length larger than $4\eps$.
This contradicts the OSE guarantee that $\|\Pi Uv\|_2^2\in (1\pm \eps)^2$, which is a range of length exactly $4\eps$.

For the dense case (i.e.~\cref{thm:main2}), the proof shares some basic intuition with the sparse case but considers all pairs from $\chi\geq 2$ columns instead of just a single pair. Here, we aim to find $\chi$ columns $X_1,\dots,X_\chi$ of $\Pi V$ so that $\sum_{i\ne j}\langle X_i,X_j\rangle \ge 4\eps\chi$. 
Now, the random $W$ does nothing (as the randomnization trick does not work for large $\chi$), but we can show that $\sum_i\|X_i\|_2^2\ge (1-\eps)^2\chi$ and then choose an appropriate unit vector $v$ such that $\|\Pi Uv\|_2^2\cdot\chi=\|\sum_i X_i\|_2^2=\sum_i\|X_i\|_2^2+\sum_{i\ne j}\langle X_i,X_j\rangle>(1+\eps)^2\chi$.

We remark that the strategy for proving \cref{thm:main} results in an upper bound of $1/(2\eps)$ for $s$ and that the strategy for proving \cref{thm:main2} yields a phase transition around $s=1/(4\eps)$ from quadratic to subquadratic dependence on $d$.

In the following sections, we shall describe our approach for the sparse case from Sections~\ref{sec: previous} to \ref{sec:our_approach} before moving on to the dense case in \cref{sec:dense_approach}. Although the limitations presented in \cref{sec: previous} and the intuition in \cref{sec:intuition} also serve the dense case, for simplicity, they are described in a manner tailored to the sparse case (i.e. $\chi=2$).

\subsection{Limitations of previous approaches}\label{sec: previous}
All the previous proofs~\cite{NN14,LL22} adopt the following strategy. Consider a row $k$ of $\Pi$ and let $S_k$ be the set of the column indices of large entries in this row, i.e., $\Pi_{k,j}\ge \sqrt\kappa$ for all $j\in S_k$, where $\kappa\in (0,1)$. As mentioned above, we seek to show that there exist two columns of $\Pi V$ with a large inner product. Here, we would like to argue that $\langle (\Pi V)_{\star,i},(\Pi V)_{\star,j}\rangle$ is large with a good small probability when $(\Pi V)_{\star,i}$ and $(\Pi V)_{\star,j}$ are uniformly distributed over the columns in $S_k$. 

To see this, we split a vector into the $k$-th coordinate and the remaining coordinates, writing
\[
\langle (\Pi V)_{\star,i},(\Pi V)_{\star,j}\rangle = (\Pi V)_{k,i}(\Pi V)_{k,j} + \langle (\Pi'_kV)_{\star,i}, (\Pi'_kV)_{\star,j}\rangle \geq \kappa + \langle (\Pi'_kV)_{\star,i}, (\Pi'_kV)_{\star,j}\rangle,
\]
where $\Pi'_k$ be the matrix obtained from $\Pi$ by zeroing out the $k$-th row. 
It suffices to prove that the remainder $X = \langle (\Pi'_kV)_{\star,i}, (\Pi'_kV)_{\star,j}\rangle$ would not be too negative to cancel $\kappa$.
For example, it can be shown that 
\begin{align}
    \Pr[ X \ge -\kappa/2]=\Omega(\kappa). \label{eqn: previous intuition}
\end{align}
Taking $\kappa = \Theta(\eps)$, this shows that two random columns of $\Pi V$ in $S_k$ have inner product $\Omega(\eps)$ with probability $\Omega(\eps)$. It then requires finding $\Omega(1/\eps)$ such pairs of columns (which may come from $S_k$ for different values of $k$) to conclude the existence of one column pair of large inner product with constant probability. A simple greedy strategy in~\cite{NN14} finds $\Omega(\sqrt{d^2/m})$ pairs, leading to a lower bound of $\Omega(\eps^2 d^2)$; a more sophisticated greedy strategy in~\cite{LL22} finds much more pairs and obtains an improved $\Omega(\eps^{\Theta(\delta)}d^2)$ lower bound.

We remark that \cref{eqn: previous intuition} cannot be improved for a general matrix $\Pi$. Furthermore, an  example was given in 
{\cite[Remark 10]{LL22}}, showing that this strategy cannot prove a lower bound beyond $\Omega(d^2)$.
The example takes $\Pi$ to be a horizontal concatenation of copies of $d^2\times d^2$ block diagonal matrix, where each diagonal block is $\sqrt{8\eps} H$ with $H$ being a Hadamard matrix of order $1/(8\eps)$. Hence, every nonzero entry of $\Pi$ is $\pm \sqrt{8\eps}$. Assume that $\delta$ is a constant. The hard instance is $\cD_1$ and the matrix $V$ has thus $d$ columns.
 The previous strategy applies \cref{eqn: previous intuition} with $\kappa=8\eps$. Now, two random columns of $\Pi V$ belong to the same $S_k$ with probability $\Theta(1/(\eps d^2))$; when this occurs, they have a large inner product with probability $\Theta(\eps)$ according to the earlier argument. Hence, two random columns of $\Pi V$ have an inner product $\Theta(\eps)$ with probability $\Theta(1/d^2)$.
On the other hand, $\Pi V$ has at most $O(d^2)$ different column pairs.
Thus, the previous strategy finds a large inner product with probability at most a small constant, which can be made smaller than $\delta$ by enlarging the size of $\Pi$ from $d^2$ to $\Theta(d^2)$. No contradiction with the subspace embedding property (\cref{eqn:SE_matrix}) can thus be derived in this manner for this $\Pi$ and hence a higher lower bound than $\Omega(d^2)$ is impossible.

In this example, however, two random columns of $\Pi V$ has a much larger inner product than $\Theta(\eps)$ whenever the inner product is nonzero; that is, $X$ is likely to be much larger than $\kappa$.
Specifically,
\[
\Pr\{X \geq 1 -8\eps\} = \Theta(\eps)
\]
	when $(\Pi'_kV)_{\star,i}$ and $(\Pi'_kV)_{\star,j}$ are drawn from the same $S_k$. This suggests that we could use a more challenging hard instance $\cD_\beta$ with a smaller $\beta$, which would be conducive to a higher lower bound. For instance, when $\beta = \Theta(\eps)$, the inner product of two columns of $\Pi V$ will be scaled down by a factor of $\Theta(\eps)$, leading to an inequality comparable with \cref{eqn: previous intuition}. But there are now potentially many more column pairs than before, making a higher lower bound more likely.

\subsection{The intuition}\label{sec:intuition}
Recall that $X=\langle (\Pi'_kV)_{\star,i}, (\Pi'_kV)_{\star,j}\rangle$ for two random $i,j\in[n]$, where $\Pi'_k$ be the matrix obtained from $\Pi$ by zeroing out the $k$-th row. 
The central technical issue is how to exploit the fact that $X$ can be much larger than $\kappa$.
We characterize the phenomenon with \cref{eqn: intuition} below. %
Note that $X\le 1$ and $\mathbb{E}[X]\ge 0$.
For all $\kappa\in(0,1)$, if $X <-3\kappa$ with probability at least $2/3$, then there is a nonnegative integer $\ell \leq \log(1/\kappa)$ such that 
\begin{align}
\Pr[X\approx 2^\ell\kappa]\approx 2^{-\ell}/\log(1/\kappa).\label{eqn: intuition}
\end{align}
This quickly follows from applying the law of total expectation: 
\begin{align*}
    &\quad\ \kappa+\sum_{\ell=0}^{\log(1/\kappa)}\mathbb{E}\left[X\mid X\in ({2^\ell\kappa},{2^{\ell+1}\kappa}]\right]\cdot\Pr\left[X\in({2^\ell\kappa},{2^{\ell+1}\kappa}]\right]\\
    &\ge \mathbb{E}[X\mid X\ge -3\kappa]\cdot\Pr[X\ge -3\kappa]\\
    &= \mathbb{E}[X]-\mathbb{E}[X\mid X<-3\kappa]\cdot\Pr[X<-3\kappa] \\ 
    &\ge 2\kappa.
\end{align*}
The formal statement and its proof can be found in \cref{lem:good inner product}. 
It is tempting to apply our new observation, \cref{eqn: intuition}, in the strategy used by \cite{NN14,LL22} as described in \cref{sec: previous} to prove a higher lower bound. However, \cref{eqn: intuition} is incompatible with the existing strategy due to a chicken-and-egg issue (see Appendix~\ref{sec:incompatibility} for explanation). Instead, we 
devise a novel approach which is drastically different from the previous greedy strategies to lower bound the number of column pairs with large inner products.

\subsection{Our approach for sparse case} \label{sec:our_approach}
To exploit \cref{eqn: intuition}, we adopt a straightforward strategy in a global manner. For description simplicity, we call a column pair \emph{good} if the pair has a large inner product.
We partition columns of $\Pi V$ into two disjoint sets of equal size $d/(2\beta)$ and show that
\begin{claim}[Informal] \label{claim:dichotomy}
One of the following two cases must hold:
\begin{enumerate}[label=(\roman*)]
	\item (good intra-partition pair) With $\Omega(1)$ probability, there exist $i,j \leq d/(2\beta)$ such that $\langle (\Pi V)_{\star,i},(\Pi V)_{\star,j}\rangle$ is large;
	\item (good inter-partition pair) With $\Omega(1)$ probability, there exist $i \leq d/(2\beta)$ and $j > d/(2\beta)$ such that $\langle (\Pi V)_{\star,i},(\Pi V)_{\star,j}\rangle$ is large.
\end{enumerate}
\end{claim}

\cref{claim:dichotomy} is proved by an incremental strategy. We shall show 
\begin{claim}[Informal] \label{claim:incremental}
 For any column set $S''\subseteq[n]$ and a uniformly random column $J\in[n]$, with a good probability, either $J\in S''$, or 
$S'_J\setminus S''$ is not small. 
\end{claim}
We first demonstrate how \cref{claim:incremental} is used to prove \cref{claim:dichotomy}.
We examine the first $d/(2\beta)$ columns of $\Pi V$ one by one. Define $S'_j$ to be the set of the columns of $\Pi$ which has a large inner product with $(\Pi V)_{\star,j}$, i.e. $S'_j
\triangleq \{i\in[n]:\langle \Pi_{\star,i},(\Pi V)_{\star,j}\rangle\text{ is large}\}$.
At the $j$-th step ($j\leq d/(2\beta)$), we take $S'' = S_1'\cup\cdots\cup S_{j-1}'$ and $J\in [n]$ to be the index such that $\Pi_{\star,J} = (\Pi V)_{\star, j}$. If $J\in S''$, we obtain a good intra-partition pair; otherwise, the size of $S''$ increases by a substantial amount. At the end of this procedure, if we have not found a good intra-partition pair, we must have increased the size of $S''$ at every step and the value $n' = |S_1'\cup\cdots\cup S'_{d/(2\beta)}|$ is thus large. Note that $n'$ is the number of columns of $\Pi$ which can form a good pair with some column in the first half of $(\Pi V)_{\star,j}$. 
Recall that the second half of $d/(2\beta)$ columns of $\Pi V$ are uniformly distributed over the $n$ columns of $\Pi$. Hence, a good inter-partition pair exists with probability at least $1-(1-n'/n)^{d/(2\beta)}$. This probability can be shown to be large, given the lower bound on $n'$. \cref{claim:dichotomy} then follows.

Now, we need to prove \cref{claim:incremental} for an arbitrary $\Pi$, but it is hardly feasible
to analyze $S'_J$ directly. Instead, we argue that it suffices to choose an appropriate value $\theta\in [\eps,1] $ with an integer $\ell\in [0,\log(1/\theta)]$ and sample a random row $k$ such that $\Pi_{k,J}\approx \sqrt{\theta}$, then lower bound $|S'_J\setminus S''|$ by lower bounding the size of $\{i\in[n]\setminus S'':\Pi_{k,i}\approx\sqrt\theta,\langle\Pi_{\star,J},\Pi_{\star,i}\rangle\approx 2^\ell\theta\}$.

\paragraph{Great collision lemma.}
To this end, we introduce the \emph{great collision lemma}, which is our core technical innovation for the sparse case.
\begin{lemma}[Great collision lemma, informal version of {\cref{lem: small inner product}}]
	Suppose that $S$ is a finite multiset of vectors in the unit $\ell_2$ ball and $\kappa\in (0,1/2]$. There exists $\ell \leq \lceil\log(1/\kappa)\rceil$ such that for every $S''\subseteq S$ of size $O(|S|/\log(1/\kappa))$, 
	\[
	\Pr_{u\sim\Unif(S)} \left[ \Abs{\{v\in S\setminus S'':\langle u,v\rangle \ge 2^\ell\kappa-2\kappa\}} = \Omega\left(\frac{|S|}{2^\ell\log(1/\kappa)}\right) \right] = \Omega\left(\frac{1}{\log(1/\kappa)}\right).
	\]
\end{lemma}
Informally speaking, the great collision lemma shows that, for any set $S$ of unit vectors, for any $\kappa>0$, there is an $\ell$ such that there are $\approx |S|^2/(2^\ell\log(1/\kappa))$ vector pairs in $S$ of inner product $\approx 2^\ell\kappa$.

The application of the great collision lemma is straightforward.
We fix universal $\theta,\ell$ by an averaging argument such that
\begin{enumerate*}[label=(\roman*)]
	\item 
	many entries of $\Pi$ have value $\approx\sqrt\theta$, and
	\item 
	for many rows $k\in[m]$, applying the great collision lemma with $\kappa=\theta/2$ on the vector set $\left\{(\Pi'_k)_{\star,j}:j\in[n],\Pi_{k,j}\approx\sqrt\theta\right\}$ returns the fixed $\ell$. %
\end{enumerate*}

We now provide more quantitative details in the earlier argument which proves Claim~\ref{claim:dichotomy} using Claim~\ref{claim:incremental}. Since a column of $\Pi V$ is uniformly distributed over all columns of $\Pi$, we denote it by $\Pi_{\star,J}$, where $J$ is uniformly distributed over $[n]$.
We sample a row $k$ such that $\Pi_{k,J}\approx\sqrt\theta$, then the joint distribution of $(k,J)$ can be rephrased as sampling first a row $k$ and then a column $J$ from the set $S_{k,\theta}\triangleq\{j\in[n]:\Pi_{k,j}\approx\sqrt\theta\}$.
If row $k$ is not one of the rows guaranteed in aforementioned (ii), we skip this column $J$, i.e. skipping a column of $\Pi V$, so from now on we suppose $k$ is such a row in (ii).
If $|S''\cap S_{k,\theta}| = \Omega(|S_{k,\theta}|/\log(1/\theta))$, then $J\in S''$ with probability $\Omega(1/\log(1/\theta))$ and a good intra-partition pair is found.
Otherwise, 
since it holds that $\Pi_{k,j}\cdot\Pi_{k,j'}\approx\theta$ for all $j,j'\in S_{k,\theta}$, due to the guarantee of (ii) and \cref{eqn: intuition}, with probability $\Omega(1/\log(1/\theta))$, there are many columns $j\in S_{k,\theta}\setminus S''$ having inner product $\approx2^\ell\theta$ with $J$, i.e. the size of $S''$ will increase by $\Omega(|S_{k,\theta}|/(2^\ell\log(1/\theta)))$.
Therefore, after examining the first $d/(2\beta)$ columns of $\Pi V$, either we find a good intra-partition pair with constant probability, or conclude that $n' = |S''|$ is large so there exists a good inter-partition pair with constant probability.

\paragraph{Proof sketch of great collision lemma.}
Let $\varphi(u,\ell,S'')\triangleq |\{v\in S\setminus S'':\langle u,v\rangle \ge 2^\ell\kappa-2\kappa\}|$.
We shall prove that, for every not-too-large set $S''\subseteq S$, there are many $u\in S$ with large $\varphi(u,\ell,S'')$.
Due to the symmetry of the inner product, intuitively the statement claims that there are many $u\in S$ with large $\varphi(u,\ell,\emptyset)$.
Our idea is to find $\log(1/\kappa)$ disjoint subsets $S_1,\dots,S_{\log(1/\kappa)}\subset S$ of equal size, where each $S_\ell$ contains only the  vectors $u\in S$ with large $\varphi(u,\ell,\emptyset)$. By the averaging principle, there is an $\ell\in[\log(1/\kappa)]$ such that the subset $S_\ell$ contains many $u$ of large $\varphi(u,\ell,\emptyset)$.
Informally speaking, the finding procedure works in the following manner:
\begin{enumerate}[label=(\arabic*)]
	\item 
	let $S_0\triangleq\emptyset$, enumerate $\ell\in \setzero{\log(1/\kappa)}$ in ascending order.
	\item
	for each $\ell$, let $S_\ell\triangleq \emptyset$. Repeat the following for $|S|/(6\log(1/\kappa))$ times: add to $S_\ell$ the $u\in S\setminus\cup_{i\le\ell} S_i$ with the largest $\varphi(u,\ell,\cup_{i\le\ell} S_i)$.
	\item 
	we claim that there must exist an $\ell$ such that there are many $u\in S\setminus \cup_{i<\ell} S_{i}$ of large $\varphi(u,\ell,\cup_{i<\ell} S_{i})$ (which is clearly larger than $\varphi(u,\ell,\emptyset)$).
	\item 
	if the assertion does not hold for all $\ell$, we end up with $\bar{S}\triangleq S\setminus\cup_{i\le\log(1/\kappa)}S_i$ of size $\frac56|S|$, and for all $\ell$, only a few pairs $u,v\in \bar{S}$ have inner product $\approx 2^\ell\kappa$.
	However, such an $\bar S$ cannot exist due to \cref{eqn: intuition}:
	for any large vector set $\bar S$, by applying \cref{eqn: intuition} with $X=\langle u,v\rangle$ for uniformly random $u,v\in \bar S$, there must be an $\ell$ such that there are many pairs $u,v\in \bar S$ of inner product $\approx 2^\ell\kappa$.
\end{enumerate}

However, this argument still does not work for an arbitrary $S''$.
To see this,
note that the threshold $\Theta(|S|/(2^\ell\log(1/\kappa)))$ for $\varphi(\cdot)$ could be much smaller than $|S''|$ and $S''$ may contain some vectors so that $\varphi(u,\ell,S'')$ is small for almost every $u$.
To complete our proof, we slightly modify the procedure as follows:
In Step (2), for each $\ell$, we let $S'_\ell$ be the subset $S''\subseteq S$ of size at most $|S|/(32\log(1/\kappa))$ which minimizes the number of $u\in S\setminus S''$ with large $\varphi(u,\ell,S'')$, then initialize $S_\ell\triangleq S'_\ell$ and add the $u$ for $|S|/(6\log(1/\eps))$ times.
The formal proof can be found in \cref{sec:great_collision_lemma}.

\subsection{Our approach for general case} \label{sec:dense_approach}
When $s$ is large, e.g. $s>9/\eps$, an average nonzero entry becomes small in absolute value and the previous strategy of finding column pairs will not work.
To see this, consider a random $\Pi$ where each column contains $s$ nonzero entries at uniformly distributed positions, each being $1/\sqrt{s}$.
If two columns of $\Pi V$ have a nonzero entry in same row, the inner product of the two columns is  $\approx 1/s<\eps/9$ only (in absolute value), which is too small to induce a contradiction using the argument in the previous section.

Instead, if we examine $\chi$ columns $X_1,\dots,X_\chi$ of $\Pi V$ simultaneously, where each distinct pair $X_i,X_j$ share a nonzero row, we have 
\[
	\Norm{\sum_iX_i}_2^2=\sum_i\|X_i\|_2^2+\sum_{i\ne j}\langle X_i,X_j\rangle\approx \chi+ \frac{\chi(\chi-1)}{s}.
\]
If there exists a unit vector $v$ such that $\Pi U v=\sum_i X_i/\sqrt\chi$, we would have $\Norm{\Pi U v}_2^2\approx 1+(\chi-1)/s$. 
Hence, if we could find such a tuple $(X_1,\dots,X_\chi)$ for a sufficiently large $\chi$, it would be possible to generate a contradiction.

First observe that the randomization trick (\cref{lemma: inner product implies anticoncentration}), which is used in the sparse case, does not work for large $\chi$, such as $\chi=\omega(1)$.
Specifically, when $\chi$ vectors are each multiplied by a random sign $\pm 1$, the sum of the pairwise inner products could be small with a high probability when $\chi$ is large. To see this, consider the example where all $\chi$ vectors are $e_1$. In this case, the sum is $O(\sqrt{\chi\log \chi})$ in absolute value with probability $1-O(1/\chi)$ by a Chernoff bound, but we need anticoncentration of magnitude $\Omega(\chi)$ to induce the contradiction for a good lower bound.
This observation tells us that we should examine the whole $\|\Pi Uv\|_2=\|\sum_i X_i\|_2/\sqrt\chi$ instead of only the inner products among a few columns.

For simplicity, we assume henceforth that all the nonzero entries of $\Pi$ have absolute values of $1/\sqrt{s}$. 
Additionally, we remove the random signs from $W$ so that every entry of $U$ is nonnegative.
It is easy to see that almost every column of $\Pi$ has an $\ell_2$-norm of $1\pm\eps$; otherwise, $\Pi$ would not an OSE for $U\sim\cD_1$.
Therefore, $\sum_i\|X_i\|_2^2=(1\pm\eps)^2\chi$, and the key to reaching a contradiction is to show that $\sum_{i\ne j}\langle X_i,X_j\rangle\notin[-4\eps\chi,4\eps\chi]$ (recall that $\|\Pi U v\|_2^2\in[(1-\eps)^2,(1+\eps)^2]$, which is a range of length $4\eps$).
A straightforward approach is finding a row $r$ such that there are $\chi>4\eps s+1$ columns $X_1,\dots,X_\chi$ in $\Pi V$ which are nonzeroes and share the same sign in row $r$.
We can then decompose $\sum_{i\ne j}\langle X_i,X_j\rangle$ into two parts: one contributed by row $r$ and the other by all the remaining rows.
The former part is exactly $\chi(\chi-1)/s > 4\eps\chi$, and the latter part can be bounded by applying our observation \cref{eqn: intuition} as in the sparse case.

However, an immediate issue arises. 
In order to leverage the fact that the sum of the inner products is large with small probability to obtain higher lower bound, we should let $U\sim\cD_\beta$ for some small $\beta$, i.e. split each coordinate of $v$ such that $\Pi V$ has $d/\beta$ columns, and each coordinate of $v$ corresponds to $1/\beta$ columns in $\Pi V$ when examining $\Pi U v$.
For $\beta<1$, we can find multiple tuples $(X_1,\dots,X_\chi)$ such that the event in \cref{eqn: intuition} happens for one of the $\chi$-tuples with good probability, but cannot find a unit vector $v$ such that $\Pi U v$ covers exactly one of desired $\chi$-tuples. 
Nevertheless, we can choose $v$ to be a normalized sum of $\chi$ canonical basis vectors such that
$\Pi U v$ covers $X_1,\dots,X_\chi$ and another $\chi/\beta-\chi$ columns.
Consider $\Norm{\Pi U v}_2^2$ via the expansion
\[
({\chi/\beta}) \Norm{\Pi U v}_2^2 = \|X\|_2^2+\|Y\|_2^2+2\langle X,Y\rangle,
\]
where $X\triangleq\sum_i X_i$ and $Y$ is the sum of the other $\chi/\beta-\chi$ columns.
We wish to show that the latter two terms do not cancel out the first term. The middle term is easy to control as $Y$ is independent of $X$ but the last term is an issue as $\langle X,Y\rangle$ depends on $X$.
A straightforward idea is to apply \cref{eqn: intuition} by including $\|Y\|_2^2+2\langle X,Y\rangle$ together with the coordinates not in row $r$ of $X_i$'s.
However, it does not work owing to a similar chicken-and-egg issue encountered when combining \cref{eqn: intuition} with previous approaches in the sparse case.

Since $Y$ is the sum of $\chi/\beta-\chi$ uniformly random columns of $\Pi$, it is easy to obtain that $\|Y\|_2=(1\pm\eps)(\chi/\beta-\chi)$ by forcing $\Pi$ to be an OSE for $U\sim\cD_{\beta'}$, where $1/\beta'=1/\beta-1$.
Next, let us focus on the inner product $\langle X,Y\rangle$, which can be written as the sum of inner products of two columns which are uniformly distributed over all columns of $\Pi$.
By the linearity of expectation, we can see that $\E[\langle X,Y\rangle]\ge 0$.
Thus it is tempting to use \cref{eqn: intuition} again to lower bound the inner product.
However, a familiar chicken-and-egg issue arises for a third time: when given a distribution of $\langle X,Y\rangle$, applying \cref{eqn: intuition} yields an $\ell$, and we must choose $\beta$ based on this $\ell$. As a result, the distribution of $\langle X,Y\rangle$ changes according to the new value of $\beta$, since $Y$ is sum of $\chi/\beta-\chi$ random columns of $\Pi$.
To see the changing of distribution, note that $\langle X,Y\rangle=\sum_i\langle X,Y_i\rangle$, where each $Y_i$ is a single random column of $\Pi$. Thus, certain values of $\langle X,Y_i\rangle$ would be barely present in the sum when $1/\beta$ is small while they are more likely to appear when $1/\beta$ is large.

We resolve this issue in a reverse way.
We show that, by including a constant factor more columns in $Y$, $\langle X,Y\rangle$ could be smaller than $-8\eps\chi$ and $\sum_{i\ne j}\langle X_i,X_j\rangle$ will not cancel $\langle X,Y\rangle$ in this case.
Specifically, if $\Pr[\langle X,Y\rangle<-\eps\chi]>1/2$, which means that $\langle X,Y\rangle$ cancels $\sum_{i\ne j}\langle X_i,X_j\rangle$, then $\Pr[\langle X,Y''\rangle<-8\eps\chi]>1/256$, where $Y''$ is the sum of $8$ independent copies of $Y$.
Therefore, $\sum_{i\ne j}\langle X_i,X_j\rangle$ will not cancel $\langle X,Y\rangle$ if $U\sim\cD_{\beta/8}$.

To summarize, we prove the lower bound with the following ingredients:
\begin{enumerate}[label=(\arabic*)]
	\item set $\chi\approx 4\eps s$, 
	\item apply \cref{eqn: intuition} to a random $\chi$-tuple of columns of $\Pi$, which share the same nonzero row, to obtain an $\ell$,
	\item set $\beta\approx 2^{-\ell}$ so that there is at least one $\chi$-tuple, which satisfies $\sum_{i\ne j}\langle X_i,X_j\rangle\approx 4\eps \chi/\beta$, with constant probability,
	\item force $\Pi$ to be an OSE for $U\sim \cD_{\beta'}$ for four different values of $\beta'$ simultaneously: 
	\begin{enumerate*}
		\item $\beta'=\beta$,
		\item $1/\beta'=1/\beta-1$,
		\item $\beta'=\beta/8$,
		\item $1/\beta'=8/\beta-1$,
	\end{enumerate*}
	so that $\sum_{i\ne j}\langle X_i,X_j\rangle+2\langle X,Y\rangle$ is either larger than $4\eps \chi/\beta$ when $U\sim\cD_{\beta}$, or less than $-4\eps \chi/(\beta/8)$ when $U\sim\cD_{\beta/8}$, and consequently obtain a contradiction.
\end{enumerate}

\section{Notation}

For $x,y\in\R$ and $\theta>0$, we write $x= y\pm\theta$ if $x\in[y-\theta,y+\theta]$. For a matrix $A$, we denote its $i$-th column vector by $A_{\star, i}$, and denote by $A'_k$ the matrix obtained by zeroing out the $k$-th row. For a finite multiset $S$, we denote by $\Unif(S)$ the uniform distribution on $S$. For a random variable $X$ and a probability distribution $\cD$, we write $X\sim \cD$ to denote that $X$ follows $\cD$.

When $d$ and $\eps$ are clear from the context, we abbreviate the event in the probability of \cref{eqn:SE_matrix} to ``$\Pi$ is a subspace embedding for $A$''.

\section{Lower Bound for Sparse Case}
\subsection{Preliminaries}
Our hard instance $\widetilde{\cD}$ is based on the same type of that in \cite{LL22}.
$\widetilde{\cD}$ is a mixture distribution of $\cD_\beta$, which is parameterized by $\beta$, on $n\times d$ matrices.
With probability $1/2$, $\widetilde{\cD}=\cD_1$; and with probability $1/2$, $\widetilde{\cD}$ is a $\cD_\beta$ for a random $1/\beta=2^\ell$ where $\ell$ is uniformly distributed over \smallmodify{$\{0,\dots,{\lceil\log s\rceil}\}$}.
We recall the definition of $\cD_\beta$ below.

\begin{definition}[Distribution $\cD_\beta$~\cite{LL22}]\label{def:D_beta}
The distribution $\cD_\beta$ ($0 < \beta\leq 1$) is defined on matrices $U\in\mathbb{R}^{n\times d}$ as follows. 
The matrix $U$ is decomposed as $U=VW$, where $V\in\mathbb{R}^{n\times d/\beta}$ and $W\in\mathbb{R}^{d/\beta\times d}$.
The matrix $V$ has i.i.d.\ columns, each $V_{\star,i}$ ($i=1,\dots,d/\beta$) is uniformly distributed among the $n$ canonical basis vectors in $\mathbb{R}^n$. 
The matrix $W$ is distributed as follows:
For each $i=1,\dots,d$, set $W_{j,i} \triangleq \sigma_{j}\sqrt{\beta}$ for $j=(i-1)/\beta+1,\dots,i/\beta$, where $\sigma_{j}\in\{-1,1\}$ are independent Rademacher variables; set the remaining entries of $W$ to zero.
\end{definition}

Following the same argument in~\cite{LL22}, we see that $U=VW$ is an isometry with probability at least $1 - \delta/(2K)$ when $n\geq Kd^2/(\beta^2\delta)$. Thus, we again assume that $K$ is large enough so that we only consider the case that $U$ is an isometry (i.e., $V$ has independent columns and the full column rank.)

\smallmodify{Our starting point is the same as~\cite{LL22}.} First, observe that if two columns of $\Pi V$ have a large inner product (in the absolute value) then $\Pi$ cannot be a subspace embedding for $U$. This is formally captured by the following lemma in~\cite{LL22}.

\begin{lemma}[{\cite[Lemma 4]{LL22}}]\label{lemma: inner product implies anticoncentration}
    Suppose that $|\langle A_{\star,p},A_{\star,q}\rangle| \ge \lambda\epsilon/\beta$ for some distinct columns $p,q$ of a matrix $A\in \mathbb{R}^{m\times d/\beta}$, where $\lambda>2$.
    Then there exists a unit vector $u\in\R^d$ such that with probability at least $1/4$ (over $W$\modify{, which is as defined in Definition~\ref{def:D_beta}})
    \[
        \Norm{AWu}_2^2 \notin [(1-\eps)^2, (1+\eps)^2].
    \]
\end{lemma}

The remaining question is to find a pair of columns with a large inner product. The basis for finding such a pair was given in   {\cite[Lemma 3]{LL22}}, which elevates the more primitive {\cite[Lemmata 8 and 9]{NN14}}. It states that given a finite collection of vectors of length at most $1+\eps$, there always exists a small fraction of vector pairs whose inner product is not too small (``small'' here means being negative  and far from $0$). The following lemma, a refinement of~\cite[Lemma 3]{LL22}, extends the inner products to large positive numbers.

\begin{lemma}[Refined from {\cite[Lemma 3]{LL22}}]\label{lem:good inner product}
    Suppose that $S$ is a finite multiset of vectors in the $\ell_2$-ball of radius $1+\eps$ and $u,v$ are independent sampled from $\Unif(S)$.
    Further suppose that $\kappa,\Delta > 0$ satisfy that $\kappa\leq \Delta\leq 1$, and $L = \lceil\log((1+\eps)^2/\Delta)\rceil$. Then one of the following must hold:
    \begin{enumerate}[label=(\roman*)]
        \item 
        \[
        \Pr[-\kappa\le \langle u,v\rangle \le \Delta ] \ge \frac{\kappa}{2\Delta (L+1)};
        \]
        \item  there exists $i\in \setzero{L-1}$ such that 
        \[
        	\Pr[2^i\Delta < \langle u,v\rangle \leq 2\cdot 2^i \Delta] \ge \frac{\kappa}{4\cdot 2^i \Delta (L+1)}.
        \]
    \end{enumerate}
\end{lemma}
\begin{proof}
    Note that
    \[
        0\le\left\|\sum_{u\in S}u\right\|_2^2 = \sum_{\substack{u,v\in S}}\langle u,v\rangle.
    \]
    Thus
    \[
        \E_{u,v} \langle u,v\rangle \ge 0.
    \]
Let $X\triangleq\langle u,v\rangle$ for simplicity.
Note that if $\Pr[X\ge -\kappa]\ge 1/2$ there is nothing to prove, we henceforth assume that $\Pr[X< -\kappa]\ge 1/2$.

As $\E[X]\ge 0$, by the law of total expectation
\[
    \E[X|X\ge 0]\cdot\Pr[X\ge 0]\ge -\E[X|X<0]\cdot\Pr[X<0]\ge \frac{\kappa}{2}.
\]
Since $X\le (1+\eps)^2$, we have that
\[
    \E[X|X\in[0,\Delta]]\cdot\Pr[X\in[0,\Delta]]
    + \sum_{i=0}^{L-1}\E[X|X\in(2^i\Delta,2^{i+1}\Delta]]\cdot\Pr[X\in(2^i\Delta,2^{i+1}\Delta]]\ge \frac{\kappa}{2}.
\]
By the pigeonhole principle, at least one of the $L+1$ terms on the LHS is at least $\kappa/(2(L+1))$.
If it is the first term in the preceding inequality, we conclude with (i); otherwise, we conclude with (ii).
\end{proof}
When applying this lemma, we always assume that it returns an $\ell\in\{-1,\dots,\lceil\log((1+\eps)^2/\Delta)\rceil\}$, i.e. the case (i) is referred to as $\ell=-1$.

\subsection{Great collision lemma}
\label{sec:great_collision_lemma}
\LinesNumbered
\begin{algorithm}[tbh]
    \caption{Finding great collision}
    \label{algo: set manipulation}
    $j \gets 1$, $S_1 \gets S$\;
    $L \gets \lceil \log((1+\eps)^2/\Delta) \rceil+1$\;
    $K \gets L\Delta/\kappa$\;
    \ForEach{$\ell \in \{-1,\dots,{L-1}\}$}
    {
        \eIf{$\ell = -1$}{
        	$p \gets -\kappa$\;
        }{
        	$p\gets 2^\ell\Delta$\;
        }
        $S'_\ell\gets S_j$\;\label{algo: eq: def of s'_ell}
        break tie arbitrarily: 
        $\displaystyle S''_\ell\gets \argmin_{S'\subseteq S_j:|S'|\le \frac{|S|}{32K}} \sum_{c\in S_j\setminus S'}\mathbf 1_{\left\{ \Abs{ \left\{c'\in S_j\setminus S':\langle c,c'\rangle \ge p \right\}}   \ge \frac{|S|}{2^{\ell+5}K}  \right\}}$\;\label{alg:line:armin}
        $S_{j+1}\gets S'_\ell\setminus S''_\ell$\; \label{alg:line:remove a chunk}
        $j\gets j+1$\;
        $i\gets 1$\;
        \While{$i\le |S|/(6K)$ and $\Abs{\left\{(c,c')\in S_j\times S_j:\langle c,c'\rangle \ge p \right\}}\ge |S|^2/(2^{\ell+3}K)$}
        {
            $\displaystyle c\gets\argmax_{c'\in S_j}\Abs{\left\{c''\in S_j:\langle c',c''\rangle \ge p \right\}}$\;\label{alg:line:armax}
            \If{$\Abs{\left\{c'\in S_j:\langle c,c'\rangle \ge p\right\}} < \kappa|S|/(2^\ell\Delta)$}
            {
                \Return{$(\ell,S'_\ell,S_j)$}\; \label{alg:line:return ell first case}
            }
                $S_{j+1}\gets S_{j}\setminus \{c\}$\;  \label{alg:line:remove a singleton}
                $j\gets j+1$\;
                $i\gets i+1$\;
        }
        \If{$i>|S|/(6K)$}
                {\Return{$(\ell,S'_\ell)$}\;\label{alg:line:return ell second case}} 
    }
\end{algorithm}

Lemma~\ref{lem:good inner product} shows that \smallmodify{for any vectors set $S$, there exists an $\ell$ such that the number of vector pairs in $S$} with inner product $\approx 2^\ell \kappa$ is at least $\approx n^2/2^\ell$. In this section, we prove something stronger: the number of vector pairs with inner product $\approx 2^\ell\kappa$ is at least $\approx n^2/2^\ell$ with high probability. This result, which we dub as the \emph{great collision lemma}, is our major technical innovation and is formally stated \smallmodify{in following \cref{lem: small inner product}}.

\modify{To find vector pairs of large inner products, we use \cref{algo: set manipulation}, which proceeds in a greedy manner.
For each $\ell$, it collects vectors that have a large inner product with many other vectors.
If there are many such vectors, the task is done (the case in Line 19). Otherwise, each remaining vector has a large inner product with not-too-many other vectors, and by Lemma~\ref{lem:good inner product} and the averaging principle, there will still be sufficient vector pairs with a large inner product (the case in Line 12).
}

\begin{lemma}[Great collision lemma]\label{lem: small inner product}
    Suppose that $S$ is a finite multiset of vectors in the $\ell_2$-ball of radius $1+\eps$, and that $0<\kappa\leq \Delta\leq 1$, $L = \lceil\log((1+\eps)^2/\Delta)\rceil+1$ and $K = L\Delta/\kappa$. Then, 
    there exists an integer $\ell\in \{-1,\dots,L-1\}$ 
    such that for every $S'\subseteq S$ with $|S'| \leq |S| / (32 K)$ it holds that
    \[
        \Pr_{c\sim\Unif(S)} \left[\Pr_{c'\sim\Unif(S\setminus S')} \left[\langle c,c'\rangle\ge p\right]\ge \frac{1}{2^{\ell+5} K} \right]  \ge \frac{3}{31 K},
    \]
    where $p\triangleq -\kappa$ if $\ell=-1$, and $p\triangleq 2^{\ell}\Delta$ if $\ell\ge 0$.
\end{lemma}
\begin{proof}
    For notational convenience, let $\lambda \triangleq 3/31$.     
Recall that $p \triangleq -\kappa$ if $\ell=-1$ and $p\triangleq 2^\ell\Delta$ if $\ell\ge 0$. 
    Consider \cref{algo: set manipulation}. We first claim that it must return an $\ell$. Line~\ref{alg:line:remove a chunk} removes in total at most $|S|/(32K)$ columns from $S_j$ and Line~\ref{alg:line:remove a singleton} at most $|S|/(6K)$ columns. Hence, if the algorithm does not return an $\ell$, we would end up with an $S_j$ such that 
    \begin{enumerate*}[label=(\roman*)]
        \item $\Abs{S_j} \geq \frac{77}{96}\Abs{S}$, and
        \item $\Abs{ \left\{(c,c')\in S_j\times S_j:\langle c,c'\rangle \ge p\right\} } < \allowbreak  |S|^2/\left(2^{\ell+3}K \right) <\allowbreak  |S_j|^2/\left(2^{\ell+2}K\right)$. %
\end{enumerate*}
But, by \cref{lem:good inner product}, (i) and (ii) cannot hold simultaneously.

We now know that the algorithm will return an $\ell$. There are two cases. %
\paragraph*{$\ell$ returned in Line~\ref{alg:line:return ell first case}.} 
The algorithm returns an $S_j$, an $S'_\ell$, together with $\ell$, which means that the ``ground'' set is $S'_\ell$ and we can find ``good'' vectors in $S_j$.

The claimed result can be rewritten as
    \[
        \sum_{c\in S} \mathbf{1}_{ \left\{ \Abs{ \{c'\in S\setminus S': \langle c,c'\rangle \geq p \} } \geq \frac{|S|}{2^{\ell+5}K} \right\} } \geq \frac{\lambda |S|}{K}
    \]
     for all $S'\subseteq S$ with $|S'|\le|S|/(32K)$.
    Recall that $S'_\ell\subseteq S$ in \cref{algo: eq: def of s'_ell}.
    To prove the claimed result, it suffices to show that for all $S'\subseteq S$ with $|S'|\le|S|/(32K)$, 
    \begin{equation} \label{eqn:ell_first_case_target}        
        P \triangleq \sum_{c\in S'_\ell\setminus S'} \mathbf{1}_{ \left\{ \Abs{ \{c'\in S'_\ell\setminus S': \langle c,c'\rangle \geq p \} } \geq \frac{|S|}{2^{\ell+5}K} \right\} } \geq \frac{\lambda |S|}{K},
    \end{equation}
and we can assume without loss of generality that $S'\subseteq S'_\ell$.
The definition of $S''_\ell$ guarantees that $S''_\ell$ is the $S'$ which minimizes 
\[
\Abs{ \left\{ c\in S'_\ell\setminus S': \Abs{ \left\{c'\in S'_\ell\setminus S': \langle c,c'\rangle \geq p \right\} }\geq \frac{|S|}{2^{\ell+5}K} \right\} },
\]
regardless of how we break the tie in \cref{alg:line:armin}.
Recall that $S_j\subseteq S'_\ell\setminus S''_\ell$.
\cref{algo: set manipulation} guarantees \smallmodify{in Line 17} that 
\[
\Abs{ \{c'\in S_j:\langle c,c'\rangle \ge p\} } < \frac{\kappa |S|}{2^\ell\Delta}, \quad \forall c \in S_j. %
\]
On the other hand, \smallmodify{combining the above equation with \cref{eqn:ell_first_case_target},}
\begin{align}
    \Abs{ \{(c,c')\in S_j\times S_j:\langle c,c'\rangle \ge p\} } 
    \leq \frac{\kappa|S|}{2^\ell\Delta} \cdot P + \frac{|S|}{2^{\ell+5} K}\cdot (|S_j|-P).\label{eqn:ell_first_case_ub}
\end{align}
\cref{algo: set manipulation} also guarantees \smallmodify{in Line 15} that 
\begin{equation}\label{eq: set manipulation case 1.1}
\Abs{ \left\{ (c,c')\in S_j\times S_j: \langle c,c'\rangle\geq p \right\} }\geq \frac{|S|^2}{2^{\ell+3}K}.
\end{equation}
Combining \smallmodify{\cref{eqn:ell_first_case_ub} with \cref{eq: set manipulation case 1.1}} yields the desired result~\cref{eqn:ell_first_case_target}.

\paragraph*{$\ell$ returned in Line~\ref{alg:line:return ell second case}.}
The algorithm returns an $S'_\ell$ together with $\ell$, which means that the ``ground'' set is $S'_\ell$ and the ``good'' vectors are the $c$'s found in \cref{alg:line:armax}.

Recall that $S'_\ell\subseteq S$ in \cref{algo: eq: def of s'_ell}.
Similarly to the previous case~\cref{eqn:ell_first_case_target}, it suffices to show that 
\begin{equation}\label{eqn:ell_second_case_target}
        \sum_{c\in S'_\ell\setminus S'} \mathbf{1}_{ \left\{ \Abs{ \{c'\in S'_\ell\setminus S': \langle c,c'\rangle \geq p \} } \geq \frac{|S|}{2^{\ell+5}K} \right\} } \geq \frac{\lambda |S|}{K}
\end{equation}
for all $S'\subseteq S$ with $|S'|\le |S|/(32K)$, 
and we can assume without loss of generality that $S'\subseteq S'_\ell$.
The definition of $S''_\ell$ guarantees that $S''_\ell$ is the $S'$ which minimizes 
\[
\Abs{ \left\{ c\in S'_\ell\setminus S': \Abs{ \left\{c'\in S'_\ell\setminus S': \langle c,c'\rangle \geq p \right\} }\geq \frac{|S|}{2^{\ell+5}K} \right\} },
\]
regardless of how we break the tie in \cref{alg:line:armin}.
After removing $S''_\ell$ from $S'_\ell$, the while-loop is executed for at least $|S|/(6K)$ times, so there are at least $|S|/(6K)$ elements $c \in S'_\ell\setminus S''_\ell$ satisfying that $\Abs{\{ c'\in S'_\ell\setminus S''_\ell: \langle c,c'\rangle\geq p\}}\geq \kappa|S|/(2^\ell\Delta)$. 
This implies that
\[
        \sum_{c\in S'_\ell\setminus S_{\ell}''} \mathbf{1}_{ \left\{ \Abs{ \{c'\in S'_\ell\setminus S''_\ell: \langle c,c'\rangle \geq p \} } \geq \frac{\kappa |S|}{2^\ell\Delta} \right\} } \geq \frac{|S|}{6K}.
\]
It follows that for a general $S'$
\begin{align*}
\sum_{c\in S'_\ell\setminus S'} \mathbf{1}_{ \left\{ \Abs{ \{c'\in S'_\ell\setminus S': \langle c,c'\rangle \geq p \} }\geq \frac{|S|}{2^{\ell+5}K} \right\} } \geq& \sum_{c\in S'_\ell\setminus S''_\ell} \mathbf{1}_{ \left\{ \Abs{ \{c'\in S'_\ell\setminus S': \langle c,c'\rangle \geq p \} }\geq \frac{|S|}{2^{\ell+5}K} \right\} } \\
\geq& \sum_{c\in S'_\ell\setminus S''_\ell} \mathbf{1}_{ \left\{ \Abs{ \{c'\in S'_\ell\setminus S': \langle c,c'\rangle \geq p \} }\geq \frac{\kappa|S|}{2^{\ell}\Delta} \right\} }  \geq \frac{|S|}{6K},
\end{align*}
whence we see that \cref{eqn:ell_second_case_target} holds automatically.
\end{proof}

\subsection{Fixing the parameters}

We shall prove a lower bound on the number of rows for an arbitrary unknown matrix $\Pi$ which is an $(\eps,\delta)$-subspace-embedding for $U = VW\sim\widetilde{\cD}$.
To analyze $\Pi V$, we shall fix some parameters of $\Pi$ and $U$ for our analysis so that 
we can apply the great collision lemma (\cref{lem: small inner product}) with appropriate parameters.

First, we %
fix the parameter $\kappa,\Delta$ in \cref{lem: small inner product} to $\theta/\log s$ and $\theta$ respectively, where $\sqrt\theta$ is the most popular magnitude among all the nonzero entries of $\Pi$.
Then, for each row $k$, we find a set $S^+_{k,\theta}$ of columns which will work perfectly with \cref{lem: small inner product}, and apply \cref{lem: small inner product} to them to obtain a good $\ell$. Now we have an $\ell = \ell(k)$ for each row $k$. We then find the most popular $\ell$, denoted by $\ell_\theta$, and focus on the corresponding rows in $\Pi$, i.e., the rows $k$ with $\ell(k) = \ell_\theta$. %
Since our overall hard distribution is a mixture of hard instances corresponding to different values of $\beta$, we also need to fix our hard instance so that the pair of $\beta$ and $\ell_\theta$ induces a high lower bound. Here, we can fix a parameter $\ell'_\theta$ based on the value of $\ell_\theta$ and assert that $\Pi$ must be a subspace embedding for $U\sim\cD_\beta$ with $\beta = 2^{-\ell'_\theta}$.

Let $L\triangleq \lceil\log s\rceil$.
Recall that our hard distribution $\widetilde{\cD}$ is $\cD_1$ with probability $1/2$ and is $\cD_{2^{-\ell}}$ for $\ell\sim \Unif(\setzero{L})$ with probability $1/2$, and that 
$\Pi$ is an $(\eps,\delta)$-subspace-embedding for $U\sim\widetilde{\cD}$.
By \cref{claim: nonzeroes}, it suffices to consider only the columns of $\Pi$ with $\ell_2$-norm of $1\pm\eps$.
\begin{lemma}[Refined from {\cite[Lemma 6]{LL22}}]\label{claim: nonzeroes}
If $\Pi\in\mathbb{R}^{m\times n}$ is an $(\epsilon,\delta)$-subspace-embedding for $U\sim\widetilde{\cD}$, then $(1-2\delta/d)$-fraction of the column vectors of $\Pi$ has $\ell_2$-norm of $1\pm\epsilon$.
\end{lemma}

\subsubsection{Finding popular entry magnitude}

\begin{lemma}\label{lem:theta}
    Suppose that $\Pi$ is a matrix with column sparsity at most $s\geq 3$. %
    Then there exists a $\theta\ge  (1-\eps)^2/s-1/s\log s$ such that 
    \begin{equation}\label{eqn:row_blocks_expectation}
        \E_{j{\sim\Unif([n])}}\left[ \sum_{k: |\Pi_{k,j}|^2\in[\theta,2\theta)} \Abs{\Pi_{k,j}}^2 \;\middle\vert\; \Norm{\Pi_{\star,j}}_2= 1\pm\eps\right]=\Omega\left(\frac{1}{\log^2 s}\right).
    \end{equation}
\end{lemma}
\begin{proof}
Let $\theta\triangleq (1-\eps)^2/s-1/s\log s$ and $I\triangleq\lceil\log_2((1+\eps)^2/\theta)\rceil$.
Since the column sparsity is $s$, it is clear that 
\[
\sum_{k: |\Pi_{k,j}|^2 \geq \theta} |\Pi_{k,j}|^2> (1 - \eps)^2 - s\theta \geq \frac{1}{\log s}.
\]
By the linearity of expectation, this implies that
\[
        \sum_{i=0}^I \E_{j{\sim\Unif([n])}}\left[ \sum_{k: |\Pi_{k,j}|^2 \in [2^i\theta,2^{i+1}\theta)} \Abs{\Pi_{k,j}}^2 \;\middle\vert\; \Norm{\Pi_{\star,j}}_2= 1\pm\eps\right] \geq \frac{1}{\log s}.
\]
Therefore there exists $i$ such that
\[
\E_{j{\sim\Unif([n])}}\left[ \sum_{k: |\Pi_{k,j}|^2 \in [2^i\theta,2^{i+1}\theta)} \Abs{\Pi_{k,j}}^2 \;\middle\vert\; \Norm{\Pi_{\star,j}}_2= 1\pm\eps\right] = \Omega\left(\frac{1}{\log^2s}\right).
\]
Let $\theta$ in the lemma statement be $ 2^i\theta$. The proof is complete.
\end{proof}

\modify{
\subsubsection{Applying the great collision lemma}
}

Let $\theta$ be as guaranteed by Lemma~\ref{lem:theta}. For each row $k$, let $S_{k,\theta}\triangleq\{j\in[n]:|\Pi_{k,j}|^2\ge\theta\text{ and }\Norm{\Pi_{\star,j}}_2= 1\pm\eps\}$ denote the set of column vectors $\Pi_{\star,j}$ such that $|\Pi_{k,j}|^2\ge \theta$.
For each row $k$, either $|\{j\in S_{k,\theta}:\Pi_{k,j}\ge\sqrt\theta\}|\ge |S_{k,\theta}|/2$ or $|\{j\in S_{k,\theta}:\Pi_{k,j}\le-\sqrt\theta\}|\ge |S_{k,\theta}|/2$.
In the former case, we let $S^+_{k,\theta}\triangleq\{j\in S_{k,\theta}:\Pi_{k,j}\ge\sqrt\theta\}$; in the latter case, we let $S^+_{k,\theta}\triangleq\{j\in S_{k,\theta}:\Pi_{k,j}\le-\sqrt\theta\}$.

\begin{lemma}\label{lem: in-row inner product}
    Suppose that $\Pi$ is a matrix with column sparsity at most $s$ ($s\geq 3$	). %
    There is an $\ell_\theta\in[-1,\lceil\log((1+\eps)^2/\theta)\rceil+1]$ and a \smallmodify{set $S_{\ell_\theta}$ of rows} such that the following holds.
    \begin{enumerate}[label=(\roman*)]
        \item 
    \[
    \E_{j{\sim\Unif([n])}}\left[\;\Abs{\left\{k\in S_{\ell_\theta}: j\in S^+_{k,\theta}\right\}} \;\middle\vert\; \Norm{\Pi_{\star,j}}_2= 1\pm\eps\right] = \Omega\left( \frac{1}{\theta\log^3 s} \right)
    \]
    \item
    For any row $k\in S_{\ell_\theta}$, let $S'_{k}\triangleq\{\Pi_{\star,j}\}_{j\in S^+_{k,\theta}}$, then for every $S'\subseteq S'_k$ with $|S'| \leq |S'_k| / (32 \lceil\log(1/\theta)\rceil\log s)$ it holds that
    \[
        \Pr_{c\sim\Unif(S'_k)} \left[\Pr_{c'\sim\Unif(S'_k\setminus S')} \left[\langle c,c'\rangle\ge p+\theta\right]= \Omega\left(\frac{1}{2^{\ell_\theta} \log^2 s}\right) \right] = \Omega\left(\frac{1}{\log^2 s}\right),
    \]
    where $p\triangleq -\theta/\log s$ if $\ell_\theta=-1$, and $p\triangleq 2^{\ell_\theta}\theta$ if $\ell_\theta\ge 0$.
    \end{enumerate}
\end{lemma}
\begin{proof}
We first show guarantee (i).
Intuitively, we shall apply the great collision lemma (\cref{lem: small inner product}) to each column, then take the majority of the returned $\ell$'s.

For each $k\in[m]$, recall that matrix $\Pi'_k$ is obtained from $\Pi$ by zeroing out the $k$-th row of $\Pi$. 
For each nonempty $S^+_{k,\theta}$, note that $\|(\Pi'_{k})_{\star,j}\|_2\le 1+\eps$ for each ${j\in S^+_{k,\theta}}$. %
Thus, for each row $k\in[m]$, we can apply \cref{lem: small inner product} to $\{(\Pi'_{k})_{\star,j}\}_{j\in S^+_{k,\theta}}$ with $\kappa=\theta/\log s, \Delta=\theta$ and obtain an $\ell_k \leq \lceil \log((1+\eps)^2/\theta)\rceil + 1$. 
Applying the averaging principle to \cref{eqn:row_blocks_expectation}, there is an $\ell_\theta$ such that 
\[
    \E_{j{\sim\Unif([n])}}\left[\;\Abs{\left\{k\in S_{\ell_\theta}: j\in S^+_{k,\theta}\right\}} \;\middle\vert\; \Norm{\Pi_{\star,j}}_2= 1\pm\eps\right] = \Omega\left( \frac{1}{\theta\log^3 s} \right),
\]
where $S_{\ell_\theta}\triangleq\{k\in[m]:\ell_k=\ell_\theta\}$. This completes the proof of item~(i).

Next we show guarantee (ii). Since $\ell_\theta$ is obtained by applying \cref{lem: small inner product} to $\{(\Pi'_{k})_{\star,j}\}_{j\in S^+_{k,\theta}}$ with $\kappa=\theta/\log s$ and $\Delta=\theta$ for each $k\in S_{\ell_\theta}$, it holds that for every $S'\subseteq S'_k$ with $|S'| \leq |S'_k| / (32 \lceil\log((1+\eps)^2/\theta) \rceil \log s)$ that
\[
    \Pr_{c\sim\Unif(S'_k)} \left[\Pr_{c'\sim\Unif(S'_k\setminus S')} \left[\langle c,c'\rangle\ge p\right]= \Omega\left(\frac{1}{2^{\ell_\theta} \log^2 s}\right) \right] = \Omega\left(\frac{1}{\log^2 s}\right),
\]
where $p\triangleq-\theta/\log s$ if $\ell_\theta=-1$ and $p\triangleq 2^{\ell_\theta}\theta$ if $\ell_\theta\ge 0$.
Guarantee (ii) follows immediately by noticing that $\Pi_{j,k}\Pi_{j',k}\ge \theta$ for all $j,j'\in S^+_{k,\theta}$, for every row $k\in S_{\ell_\theta}$.
\end{proof}

Let $S_\theta\triangleq \{(k,j)\in[m]\times[n]:j\in S^+_{k,\theta} \text{ and } k \in S_{\ell_\theta}\}$.
We say an entry $(i,j)$ of $\Pi$ is \emph{good} if $(i,j)\in S_\theta$.
\cref{lem: in-row inner product}(i) can be rephrased as
\begin{align}
    \E_{j{\sim\Unif([n])}}\left[ \Abs{\{k\in[m]:(k,j)\in S_\theta\}} \;\middle\vert\; \Norm{\Pi_{\star,j}}_2= 1\pm\eps \right]= \Omega\left(\frac{1}{\theta\log^3 s}\right).\label{eq: key_general_sparsity}
\end{align}

The following corollary follows easily from combining \cref{eq: key_general_sparsity} and \cref{claim: nonzeroes}.
\begin{corollary}\label{claim: entries}
If $\Pi\in\mathbb{R}^{m\times n}$ is an OSE for $U\sim\widetilde{\cD}$, then $\Pi$ has at least $\Omega(n/(\theta\log^3 s))$ good entries.
\end{corollary}
\modify{
\subsubsection{Fixing the hard instance}
}

Recall $L\triangleq \lceil\log s\rceil$.
Also recall that \smallmodify{our hard instance $\widetilde{\cD}$ is a mixture of $\cD_\beta$ with different values of $
\beta$} and $\Pi$ is an $(\eps,\delta)$-subspace-embedding for $U\sim\widetilde{\cD}$.
Recall that we assume $n\ge Kd^2/(\eps^2\delta)$ for large enough constant $K$ so that $U\sim\widetilde{\cD}$ is an isometry with probability $1-\delta/(2K)$.
By Markov's inequality, for $(1-\gamma)$-fraction of $\ell\in \setzero{L}$, $\Pi$ is an $(\eps,2\delta/\gamma)$-subspace-embedding for $U\sim\cD_{2^{-\ell}}$, where $\gamma\triangleq K \delta < 1/2$, so that $2\delta/\gamma=2/K$ is a small constant.
Hence, for each $\ell\in \setzero{L}$, there exists an $\ell'\in[\max\{0,(1-\gamma)L-\ell\},L-\ell]$
 such that
\begin{align}
    \begin{cases}
    2^{-\ell-\ell'}\in[ 2^{-L} , (2^{-L})^{1-\gamma}],\text{ and }\label{eq: ell}\\
\text{$\Pi$ is an $(\eps,2\delta/\gamma)$-subspace-embedding for $U\sim\cD_{2^{-\ell'}}$.}
    \end{cases}
\end{align}
Now, if $\ell_\theta<0$, we let $\ell'_\theta=0$; otherwise let $\ell'_\theta$ be the $\ell'$ as guaranteed by \cref{eq: ell} with $\ell=L-(\ell_\theta+\log(\theta/\eps))$.
We shall apply $\Pi$ to $U\sim\cD_{2^{-\ell'_\theta}}$ and show that there is a unit vector $v$ such that $\|\Pi U v\|_2\ne 1\pm\eps$ with probability larger than $2\delta/\gamma$.
Note that
\begin{align}
    2^{\ell'_\theta}=\Omega( \theta 2^{\ell_\theta}/\eps s^{\gamma}).\label{eq: ell' and ell}
\end{align}

\subsection{The lower bound}
We prove the lower bound in this section.
Recall that we assume
\begin{align}
    3\leq s\le \frac{(1-\eps)^2+(2\eps-\eps^2)/\log s+1/\log^2s }{2\eps}.\label{eq: s upper bound}
\end{align}

Let $(C_1,C_2,\dots,C_{d'})\in[n]^{d'}$ be the columns among the $d'\triangleq2^{\ell'_\theta}d$ columns chosen from $[n]$ by $V$ in the order they are sampled.
We shall divide $\{C_i\}_{i\in [d']}$ into $\{C_i\}_{i\le d'/2}$ and $\{C_i\}_{i>d'/2}$, and prove that the following set 
\begin{equation}\label{eq: target set}
    \left\{c\in[n]:\exists i\le d'/2, \langle c,C_i\rangle \ge 2^{\ell_\theta}\theta\right\}.
\end{equation}
is large with high probability.
\modify{
    To this end, we shall examine columns $\Pi_{\star, C_i}$ one by one and maintain a growing set $S'\subseteq [n]$ consisting of the columns which have a large inner product with one of the examined columns $\Pi _{\star, C_1},\dots,\Pi _{\star, C_i}$.
}

    We wish to apply \cref{lem: small inner product} to lower bound the increment of $S'$ after examining a new $\Pi_{\star,C_i}$, but \cref{lem: small inner product} only works in the following scenario (as the inner product at least $-\kappa$ when $\ell=-1$):
    There is a row $r$ and a set of columns $S_r$ such that
    \begin{enumerate*}[label=(\roman*)]
        \item $\Pi_{r,i}\cdot \Pi_{r,j}\ge 2\eps+\kappa$ for any columns $i,j\in S_r$, and 
        \item the column $c$ in \cref{lem: small inner product} is sampled from $S_r$ uniformly at random.
    \end{enumerate*}
    Then we apply \cref{lem: small inner product} to the set of column vectors $S_r$ with row $r$ removed, to find $S'_r\subseteq S_r$ such that $\langle c',c\rangle\ge p$ for any $c'\in S'_r$ after removing row $r$. This implies that for any $c'\in S'_r$, $\langle c',c\rangle \ge p+2\eps+\kappa\ge 2\eps$ because of the aforesaid (i).

    To apply \cref{lem: small inner product} (recall the aforesaid (i) and (ii)), we employ rejection sampling and double counting in Section 7.1, which allows for viewing a random column $C_i$ as being sampled from the aforesaid $S_r$ with a sampled row $r$. Then, in Section 7.2, we combine everything and follow the proof idea in Section 2.3 to derive the desired lower bound.

\subsubsection{Sampling entries}
Let $s'_{\max}\triangleq\max_{j\in[n]}|\{k\in[m]:(k,j)\in S_\theta\}|$.
To lower bound the size of the set in \cref{eq: target set}, every time we examine a column $C_i\sim\Unif([n])$, we choose a row $k$ at random by the following procedure:
\begin{enumerate}
    \item let $s'$ be the columns sparsity of $C_i$;
    \item sample a random number $\alpha\in[0,1]$; 
    \item if $\alpha\ge s'/s'_{\max}$, discard $C_i$; otherwise, choose a row $k$ among the set \modify{$\{k'\in[m]:(k',C_i)\in S_\theta\}$} uniformly at random.
\end{enumerate}
If we do not discard $C_i$, we include the columns $j$ with properties $(k,j)\in S_\theta$ and $\langle \Pi_{\star,j},\Pi_{\star,C_i}\rangle\ge 2^{\ell_\theta}\theta$ to the set in \cref{eq: target set}.

By a double counting argument, it is easy to see that, if we do not discard $C_i$, then \modify{$(k,C_i)$} is uniformly \smallmodify{distributed} at random over all the $(k,j)\in S_\theta$.

\smallmodify{We call a column \emph{good} if the column is sampled by $V$ and is not discarded by the procedure above.}
Let $g$ be the number of good columns.
By combining \cref{claim: entries} with the fact $s'_{\max}\le 2/\theta$, any single column \smallmodify{$(\Pi V)_i$} is discarded with probability at most $O(1/\log^3 s)$.
By a Chernoff bound, $g= \Omega(d'/\log^3 s)$ with high probability.
Consequently we assume $g\ge d'/\log^3 s$ in the remainder of this paper.
Let $(C'_1,C'_2,\dots,C'_{g})\in[n]^{g}$ be the remaining columns, and $(R'_1,R'_2,\dots,R'_{g})\in[m]^{g}$ the rows correspond to the remaining columns.
Note that for any $i\in[g]$, $(R'_i,C'_i)$ is uniformly at random over all the $(k,j)\in S_\theta$.
The uniform distribution can be rephrased as following:
\begin{enumerate}
    \item 
    Let $R'_i=r$ with probability proportional to $|\{j\in[n]:(r,j)\in S_\theta\}|$;
    \item 
    Let $C'_i$ be the column of a nonzero entry in $\{j\in[n]:(r,j)\in S_\theta\}$ uniformly at random.
\end{enumerate}

\subsubsection{Finding a large inner product}
We use \cref{algo: columns} to find a pair of columns which has a large inner product. The algorithm examines \smallmodify{the columns} from first half of the good columns one by one, and checks if one of them can be used to form a large inner product with one of the examined column. If this fails, we shall show that one of the remaining half good columns can be used to form a large inner product with one of the examined columns with at least a (small) constant probability. The detailed analysis is given below.

\begin{algorithm}[tb]
    \caption{Collecting columns}
    \label{algo: columns}
    Let $g$ be the number of good columns which are sampled by $V$ and are not discarded by us\;
    Let $(C'_1,C'_2,\dots,C'_{g})\in[n]^{g}$ be the good columns, and $(R'_1,R'_2,\dots,R'_{g})\in[m]^{g}$ the rows correspond to the good columns chosen by us\;
    $S'\gets\emptyset$, $i\gets 1$\;
    $p \gets -\theta/\log s$ if $\ell_\theta=-1$ and $p\gets 2^{\ell_\theta}\theta$ if $\ell_\theta\ge 0$\;
    \While{$i\le g/2$}
    {
        \If{$C'_i\in S'$}
        {
            \Return{Collision}\;
        }
        $S'\gets S'\cup \left\{c\in [n]:(R'_i,c)\in S_\theta \text{ and } \langle \Pi_{\star,C'_i},\Pi_{\star,c}\rangle \ge p+\theta\right\}$\;
    }
\end{algorithm}

Let $S_{R'_i}\triangleq \{c\in[n]:(R'_i,c)\in S_\theta\}$ be the set column indices corresponding to entries of $S_\theta$ in row $R'_i$.
If $|S'\cap S_{R'_i}|\ge |S_{R'_i}|/(32\log(1/\theta)\log s)$, then \modify{there is a $j<i$ such that $\inner{\Pi_{\star,C'_j}}{\Pi_{\star, C'_i}}$} is large with probability at least $1/(32\log(1/\theta)\log s)$ (recall that $C'_i$ is uniformly random over $S_{R'_i}$). 
Otherwise, $|S'\cap S_{R'_i}|\le |S_{R'_i}|/(32\log(1/\theta)\log s)$. Recall the definition of a good entry and the fact that $\ell_{R_i'}$ is obtained by applying \cref{lem: small inner product} to the column set \modify{$\{(\Pi'_{R'_i})_{\star,c}\}_{c\in S_{R'_i}}$}. It follows that $|S'|$ will increase by $\Omega(|S_{R'_i}|/(2^{{\ell_\theta}}\log^2 s))$ with probability at least $\Omega(1/\log^2 s)$.

Recall that $S_\theta$ has at least $\Omega(n/(\theta\log^3 s))$ elements by \cref{claim: entries}.
Therefore, with probability at least $1/2$, $R'_i$ is a row which has at least $\Omega(n/(m\theta\log^3 s))$ entries in $S_\theta$.
Hence, with high probability, in the while-loop, at least $g/4$ times we have a row $R'_i$ which has at least $\Omega(n/(m\theta\log^3 s))$ entries in $S_\theta$.
Now we consider only such $i$.
If $|S'\cap S_{R'_i}|\ge |S_{R'_i}|/(32\log(1/\theta)\log s)$ happens \modify{at least} $g/4$ times, we find a pair of columns with a large inner product with probability at least
\begin{align*}
&\quad\ 1 - \left(1 - \frac{1}{32\log(1/\theta)\log s}\right)^{g/4} \\
&\geq 1 - \exp\left( -\frac{g}{128\log(1/\theta)\log s}\right) \\
&\geq 1 - \exp\left( - \Omega\left(\frac{d'}{\log^4 s}\right)\right)   &&{\text{(since $g\ge \frac{d'}{18\log^2 s}$ and $1/\theta\leq 2s$)}} \\
&\geq 1 - \exp\left( -\Omega\left(\frac{2^{\ell'_\theta} d}{\log^4 s} \right)\right) \\
&\geq \frac{2}{3}. &&\text{(since \cref{eq: ell' and ell}, $d\geq 1/\eps^2$ and $s\le 1/2\eps$)}
\end{align*}
 Otherwise, $|S'\cap S_{R'_i}|\le |S_{R'_i}|/(32\log(1/\theta)\log s)$ happens at least $g/4$ times.
Combining \cref{lem: in-row inner product}(ii)
and the fact that $|S_{R'_i}|=\Omega(n/(m\theta\log^3 s))$ gives that $|S'|$ increases by at least $\Omega(n/(2^{{\ell_\theta}}m\theta\log^5 s))$ with probability $\Omega(1/\log^2 s)$ each time. Hence, by a Chernoff bound, with probability at least
\[
1 - \exp\left(-\Omega\left(\frac{g}{\log^2 s}\right) \right) =1-o(1)
\]
we have $|S'|$ increases at least $\Omega(g/\log^2 s)$ times and thus
\begin{equation} \label{eqn:good size of S'}
|S'| = \Omega\left(\frac{g n}{2^{{\ell_\theta}}m\theta\log^7 s}\right).
\end{equation}
To summarize, when \cref{algo: columns} terminates, either (i) we have found a pair of columns with a large inner product with probability at least $1/2$, or (ii) we have with probability at least $1/2$ a large $S'$ satisfying \cref{eqn:good size of S'}.

In case (ii), there are still at least $g/2$ columns which have not been examined. By the forming of $S'$, each unexamined column $c$ has a large inner product with some examined column $C'_i$ ($i\leq g/2$) with probability at least $\Pr_{c\sim\Unif(G)}[c\in S'] = |S'|/n = \Omega(g/(2^{{\ell_\theta}}m\theta\log^7 s))$. Hence, there exists a pair of columns with a large inner product with probability at least 
\begin{align}
    &1-\left(1-\Omega\left(\frac{g}{2^{{\ell_\theta}}m\theta\log^7 s}\right)\right)^{g/2}\geq 1 - \exp\left( -\Omega\left(\frac{g^2}{2^{{\ell_\theta}}m\theta\log^7 s}\right) \right).\label{eq: final probability}
\end{align}
Recall that $g= \Omega(d'/\log^3 s)$, $d'=2^{\ell'_{\theta}}d$, and \cref{eq: ell' and ell}. The $\Omega(\cdot)$ quantity in the exponent on the right-hand side of~\cref{eq: final probability} is thus
\begin{align*}
   \Omega\left(\frac{d'^2}{2^{{\ell_\theta}}m\theta\log^{13} s}\right) 
   &= \Omega\left(\frac{ 2^{2{\ell_\theta}}\theta^2d^2}{2^{{\ell_\theta}}m\theta\log^{13} s\cdot \eps^{2}s^{2\gamma}}\right)  \\
   &= \Omega\left(\frac{2^{{\ell_\theta}}\theta d^2}{m\log^{13} s^{1+2\gamma}\cdot\eps^{2}}\right)  = \Omega\left(\frac{d^2}{ms^{1+2\gamma}\eps^{2}\log^{13} s}\right).
\end{align*}
If $m \leq c_0 d^2/(s^{1+2\gamma}\eps^{2} \cdot \log^{13} s)$ for some constant $c_0$, we would have the probability in \cref{eq: final probability} at least $2/3$. 
Combining all the ``with high probability'' events and both cases (i) and (ii), we see that with probability at least $1/2$, there exists a pair of columns of \modify{$\Pi V$} which has inner product at least $p+\theta$ in absolute value. 

If $\ell_\theta=-1$, we have $\ell'_\theta=0$.
In this case, we found a pair of column vectors whose inner product is at least $p+\theta$ in absolute value.
Recall that $p=-\theta/\log s$ and $\theta\ge((1-\eps)^2-1/\log s)/s$ by \cref{lem:theta}, then $p+\theta>2\eps$ by \cref{eq: s upper bound}, i.e. our assumption on $s$. %
By applying \cref{lemma: inner product implies anticoncentration} with $1/\beta=1$, $\Pi$ is a subspace embedding for \smallmodify{$U=VW\sim\cD_{2^{-\ell_\theta'}}$} with probability at most $1/8$, this contradicts \cref{eq: ell} since $1-2\delta/\gamma \ge 1 - 2/K$. %

If $\ell_\theta\ge 0$, we found a pair of column vectors whose inner product is at least $p+\theta$ in absolute value.
Recall that $p=2^{\ell_\theta}\theta$ and $\theta\ge((1-\eps)^2-1/\log s)/s$ by \cref{lem:theta}, then $p+\theta>2\eps 2^{\ell_\theta}$ by \cref{eq: s upper bound}, i.e. our assumption on $s$. %
By applying \cref{lemma: inner product implies anticoncentration} with $1/\beta=2^{\ell'_\theta}$, $\Pi$ is a subspace embedding for \smallmodify{$U=VW\sim\cD_{2^{-\ell_\theta'}}$} with probability at most $1/8$, this contradicts \cref{eq: ell} since $1-2\delta/\gamma \ge 1 - 2/K$. %

We can therefore conclude it must hold that $m=\Omega(d^2/(s^{1+2\gamma}\eps^{2}\cdot\log^{13} s))$. This completes the proof of Theorem~\ref{thm:main} by applying Yao's minimax principle.

\section{Lower Bound for General Case}\label{sec: genercal}

Our goal in this section is to prove Theorem~\ref{thm:main2}, for OSEs with a general column sparsity $s$. Recall that $\eps$ is at most a sufficiently small constant.

We shall modify the definition of the hard instance $\cD_\beta$ by changing Rademacher variables to $+1$s in $U\sim \cD_\beta$. We also define an additional hard instance $\cD_\beta' = \cD_{\beta/(\beta+1)}$ (under the new definition of $\cD_\beta$) so that $U\sim \cD'_\beta$ contains exactly $(1+1/\beta)$ nonzero entries per column. For notation convenience, when we say $1/\beta = 0$, it is understood that $\cD'_\beta$ is $\cD_1$.

\begin{definition}[Modified definition of $\cD_\beta$]
The distribution $\cD_\beta$ ($0 < \beta\leq 1$) is defined on matrices $U\in\mathbb{R}^{n\times d}$ as follows. 
The matrix $U$ is decomposed as $U=VW$, where $V\in\mathbb{R}^{n\times d/\beta}$ and $W\in\mathbb{R}^{d/\beta\times d}$.
The matrix $V$ has i.i.d.\ columns, each $V_{\star,i}$ ($i=1,\dots,d/\beta$) is uniformly distributed among the $n$ canonical basis vectors in $\mathbb{R}^n$. 
The matrix $W$ is distributed as follows:
For each $i=1,\dots,d$ and $j=(i-1)/\beta+1,\dots,i/\beta$, set $W_{j,i} \triangleq \sqrt{\beta}$; set the remaining entries of $W$ to zero.
\end{definition}

\paragraph*{Proof strategy.}
First, assume that 
\begin{enumerate*}[label=(\roman{*})]
    \item $\Pi$ is an OSE for $U\sim \cD'_\beta$ with $1/\beta\in\{0,\dots,s\}$, and 
    \item every nonzero entry of $\Pi$ is $1/\sqrt s$ in absolute value, and
    \item every column of $\Pi$ has $s$ nonzero entries. 
\end{enumerate*}
(We shall remove these assumptions in \cref{sec: removing assumption in denser case}.)
We break the matrix into $\approx m/s$ submatrices of dimension $\approx m\times ns/m$. 
Recall that each column of $\Pi U$ is the aggregation of $1/\beta+1$ columns of $\Pi$, hence we say each column of $\Pi U$ ``selects'' $1/\beta+1$ columns of $\Pi$.
It is not difficult to see that there are some submatrices $\Pi_i$ such that $\chi\approx\eps s$ columns of $\Pi U$ land in $\Pi_i$ with good probability.
For each such submatrix $\Pi_i$, suppose that the set of nonzero coordinates of $U e_1', U e_2',\dots, U e_\chi'$ covers the columns of $\Pi$ landing in $\Pi_i$, where $e_1',\dots,e_\chi'$ are $\chi$ canonical basis vectors in $\R^{d}$. For notation simplicity, we assume that $e_q' = e_q$ for $q=1,\dots,\chi$ and examine $\|\Pi U (e_1+\dots+e_{\chi})/\sqrt{\chi}\|_2$.

Let $X$ denote the sum of the $\chi$ columns selected by $U(e_1+\dots+e_{\chi})$ in $\Pi_i$ and $Y$ denote the sum of the remaining $\chi/\beta$ columns selected by $U (e_1+\dots+e_{\chi})$.
Then 
\[
    \|\Pi U (e_1+\dots +e_{\chi})/\sqrt{\chi}\|_2^2=\|X\|_2^2/(\chi/\beta+\chi)+2\langle X,Y\rangle/(\chi/\beta+\chi)+\|Y\|_2^2/(\chi/\beta+\chi).
\]
Recall that $\Pi$ is an OSE for $U\sim\cD'_\beta$ with every $1/\beta\in\{0,\dots,1/\eps\}$ by our assumption.
Note that $Y$ is the sum of $\chi/\beta$ uniformly random columns of $\Pi$.
We apply the following trivial claim to show that
$\|Y\|_2^2\ge (1-\eps)^2(\chi/\beta)$ (case $U\sim \cD_{\beta}$) with probability $1-\delta$.
\begin{claim}\label{claim: length of Y}
    If $\Pi$ is an OSE for $U\sim\cD_\beta$, then $\Pr[\|Y\|_2=(1\pm\eps)(\chi/\beta)]\ge 1-\delta$, where $Y$ is the sum of $\chi/\beta$ random columns of $\Pi$.
\end{claim}
Thus it suffices to show that $\|X\|_2^2\approx \chi+4\eps(\chi/\beta+\chi) $, and either
\begin{enumerate*}[label=(\roman{*})]
    \item $\exists\beta,\langle X,Y\rangle\ge -2\eps (\chi/\beta+\chi)$, or
    \item $\exists\beta,\langle X,Y\rangle\le -8\eps (\chi/\beta+\chi)$.
\end{enumerate*}

\subsection{Preliminaries}
We first state a simple proposition which will be needed later. It can be proven easily by Chebyshev's inequality and thus a formal proof is postponed to Appendix~\ref{sec:sum of negatively correlated}.
\begin{claim}\label{claim: sum of negatively correlated}
    Let $(X_1,\dots,X_n)\in \{0,a_1\}\times\dots\times \{0,a_n\}$ be a tuple of bivalued random variables.
    If $X_i$'s are negatively correlated and $a_i\le 1$ for all $i$, then $\Pr[\sum_i X_i\in [\mu/2,3\mu/2]]\ge 1-4/\mu$, where $\mu\triangleq \E[\sum_i X_i]$.
\end{claim}

We shall apply the following proposition to show that $\|X\|_2^2$ may be large. The corollary is a trivial extension of \cref{lem:good inner product} and its proof is postponed to Appendix~\ref{sec:k good inner product}.

\begin{corollary}\label{lem:k good inner product}
    Suppose that $S$ is a finite multiset of vectors in the $\ell_2$-ball of radius $1+\eps$ and $v_1,v_2,\dots,v_k$ are $k$ vectors independent sampled from $\Unif(S)$. 
    Further suppose that $\kappa,\Delta > 0$ satisfy that $\kappa\leq \Delta\leq 1$, and $L = \lceil\log((1+\eps)^2/\Delta)\rceil$. Then one of the following must hold:
    \begin{enumerate}[label=(\roman*)]
        \item \[
        \Pr \left[-k^2\kappa\le \sum_{\substack{i, j\in[k]\\ i\neq j}}\langle v_i,v_j\rangle \le k^2\Delta \right] \ge \frac{\kappa}{2\Delta(L+1)};
        \]
        \item  there exists $i\in\{0,\dots,L -1\}$ such that 
        \[
            \Pr\left[2^ik^2\Delta < \sum_{\substack{i, j\in[k]\\ i\neq j}}\langle v_i,v_j\rangle \leq 2\cdot 2^i k^2\Delta\right] \ge \frac{\kappa}{4\cdot 2^i \Delta (L+1)}.
        \]
    \end{enumerate}
\end{corollary}
When applying this corollary, we always assume that it returns an $\ell\in\{-1,\dots,L\}$, i.e. the case (i) is referred to as $\ell=-1$.

\subsection{A lower bound under assumption}
In this subsection we assume that every nonzero entry of $\Pi$ is $\pm 1/\sqrt{s}$, each column contains exactly $s$ nonzeroes, and $\Pi$ is an OSE for every $\cD'_\beta$ with $1/\beta\in \{0,\dots,1/\eps\}$.
We shall prove the following lower bound in this subsection with these assumptions on $\Pi$ and remove the assumptions in \cref{sec: removing assumption in denser case}.

\begin{theorem}
    Suppose that
    \begin{enumerate*}[label=(\roman*)]
    \item
    every nonzero entry of $\Pi$ is $\pm 1/\sqrt{s}$,
    \item
    each column contains exactly $s\in(2,O(\eps^{-1}\log d))$ nonzeroes, and 
    \item 
    $\Pi$ is an OSE for $U$ of distribution $\cD'_\beta$ for every $1/\beta\in\{0,\dots,1/\eps\}$.
    \end{enumerate*}
    Then $\Pi$ must have at least
    \[
    m=\Omega\left(\frac{s^{-1/(\chi-1)}}{\log^{2/{(\chi-1)}}s}\cdot\left(\frac{d}{\eps}\right)^{1+1/(\chi-1)}\right)
    \]
    rows, where $\chi = \lfloor 4\eps s\rfloor+4$.
\end{theorem}
Note that for $s=\Omega(\log(1/\eps)/\eps)\cap O((\log d)/\eps)$, the lower bound becomes
\[
m = \Omega\left( \left(\frac{d}{\eps}\right)^{1+\frac{1}{\lfloor 4\eps s\rfloor+4}}\cdot \left(\frac{\log^3(1/\eps)}{\eps}\right)^{\Theta\left(\frac{1}{\log(1/\eps)}\right)}\right) = \Omega\left(\left(\frac{d}{\eps}\right)^{1+\frac{1}{\lfloor4\eps s\rfloor+4}}\right).
\]

From now on, we let $\chi\triangleq \lfloor 4\eps s\rfloor+4$.

\subsubsection{Partitioning OSE matrix}
We partition submatrices according to the following claim.
\begin{claim}\label{claim: partition bucket}
    The matrix $\Pi\in\mathbb R^{m\times n}$ can be partitioned into $4m/s$ (combinatorial) submatrices $\Pi_1,\dots,\Pi_{4m/s}\in\mathbb R^{m\times n/(4m/s)}$ such that for each $\Pi_i$ with $i\le 2m/s$ there is a row $r$ such that every entry in row $r$ is nonzero and has the same sign.
\end{claim}
The proof is easy.
Whenever a matrix has at least $n/2$ columns, by the averaging principle, there must be a row which contains at least $ns/4m$ entries of $1/\sqrt s$ or at least $ns/4m$ entries of $-1/\sqrt s$.
We can just partition $\Pi$ in a greedy manner.

If we choose $d/\beta$ columns of $\Pi$ uniformly at random, then there will be $\approx(m/s)\binom{d/\beta}{\chi}(s/m)^\chi$ submatrices which receive at least $\chi$ of the $d/\beta$ random columns.
We are going to examine each of such submatrices from which at least $\chi$ columns are chosen.
For each of such submatrices $\Pi_i$, we can choose $\chi$ canonical basis vectors $e'_1,\dots,e'_\chi$ so that $U(e'_1+\dots+e'_\chi)$ covers at least $\chi$ columns in submatrix $\Pi_i$.
Let $X$ denote the sum of the $\chi$ columns in $\Pi_i$ and $Y$ the sum of the remaining columns outside of $\Pi_i$.
Then 
\[
    \|\Pi U (e_1+\dots +e_{\chi})/\sqrt{\chi}\|_2^2=\|X\|_2^2/(\chi/\beta+\chi)+2\langle X,Y\rangle/(\chi/\beta+\chi)+\|Y\|_2^2/(\chi/\beta+\chi).
\]
Recall that $Y$ is simply the sum of $1/\beta$ columns of $\Pi$, then $\|Y\|_2=(\chi/\beta)(1\pm\eps)$ with probability $1-\delta$ by \cref{claim: length of Y}.
Thus, it suffices to focus on the first two terms.

For each $i\le 2m/s$, let $\Pi'_i$ denote the set of columns obtained from the columns in $\Pi_i$ by removing row $r$ of $\Pi_i$.
We apply \cref{lem:k good inner product} to each of $\Pi'_i$ for $i\le 2m/s$ with $k=\chi$, $\kappa=1/(s\log s)$, $\Delta=1/s$ and obtain an $\ell$ for each $\Pi_i'$. Let $\ell_1$ be the majority of these $2m/s$ values of $\ell$ and $\cB\subseteq[2m/s]$ the index set of $i$'s such that \cref{lem:k good inner product} returns $\ell_1$ when applied to $\Pi'_i$. %
Note that $|\cB|= \Omega(m/(s\cdot\log^2 s))$ with constant probability.
Let $X$ be a random variable which is obtained by sampling $i\in\cB$ uniformly at random then summing over $\chi$ uniformly random columns in $\Pi_i$.
Let $\cA$ be the event that $$\chi^2(1/s-1/s\log s)\le \|X\|_2^2-(1-1/s)\chi \le 2\chi^2/s$$ if $\ell_1=-1$, or $$(2^{\ell_1}+1)\chi^2/s < \|X\|_2^2-(1-1/s)\chi \leq (2^{\ell_1+1}+1) \chi^2/s$$ if $\ell_1\ge 0$.
Note that $\Pr[\cA]=\Omega(1/(2^{\ell_1}\log^2 s))$ 
if $i\in \cB$ by \cref{lem:k good inner product} and \cref{claim: partition bucket}.

\subsubsection{Obtaining lower bound via anticoncentration}
Let $\cA_i$ be the event that $U$ chooses at least $\chi$ columns in $\Pi_i$, $Y'$ a uniformly random column over $\Pi$ and $\bar Y^\ell$ the sum of $\chi 2^\ell$ independent copies of $Y'$.
We are going to invoke the following lemma to complete the proof.
\begin{lemma}\label{lem: distortion implies lower bound}
    Assume that $s=O(\log d/\eps)$ and $s\ge 3,d\ge 1/\eps^9$.
If there exists $\beta = 2^{-\ell}$ such that  $\Pi$ is an OSE for $\cD'_\beta$ and
\begin{equation}\label{eqn:sum of p_i}
\E_{i\sim\Unif(\cB)}\left[ \Pr\left[ \left.\Norm{X+\bar Y^{\ell}}_2^2 \notin (1\pm\eps)^2(\chi/\beta+\chi) \right\vert \cA_i \right] \right] = \Omega\left(\frac 1 {2^{\ell}\log^2 s}\right),
\end{equation}
then it must hold that 
\[
    m=\Omega\left(\frac{2^{\ell}\cdot s^{-1/(\chi -1)}}{\log^{2/{(\chi-1)}} s}\cdot\left(\frac{d}{\eps}\right)^{\chi/(\chi-1)}\right).
\]
\end{lemma}
\begin{proof}
Let $\cB'\subseteq\cB$ denote the set of submatrices which receive at least $\chi$ columns from $U$.
Let $X_i$ be indicator random variable for $i\in \cB$ having at least $\chi$ columns from $U$.
Note that $X_i$'s are negatively correlated, by \cref{claim: sum of negatively correlated}, with constant probability
\begin{align}
    |\cB'|=\Theta\left( |\cB|\cdot\binom{d/\beta+d}{\chi}\left(\frac{s}{m}\right)^\chi\right).
\end{align}
If the RHS is $O(2^\ell\log^2s)$, we are done. In the remainder of the proof, we thus assume that $|\cB'|=\Omega(2^\ell\log^2s)$ with constant probability. Choose a subset $\cB''\subseteq \cB'$ of size $\Theta(2^\ell\log^2s)$ uniformly at random over $\cB'$.
Conditioned on $|\cB''|=k$, it is easy to see that $\cB''$ is uniformly distributed over all $k$-tuples of $[2m/s]$ 
without replacement by the symmetry of distribution $\widetilde{\cD}$.

Each submatrix $\Pi_i$ with $i\in\cB''$ may receive more than $\chi$ columns from $U$.
Thus, we choose $\chi$ columns $\{C_{i,j}\}_{j\in[\chi]}$ among these columns uniformly at random.
Note that $\{C_{i,j}\}_{j\in[\chi]}$ is distributed uniformly at random over $(\mathcal C_i)^\chi$, where $\mathcal C_i$ is the set of all columns of submatrix $\Pi_i$.
For each $\{C_{i,j}\}_{j\in[\chi]}$ with $i\in\cB''$, let $\{e_{i,j}\}_{j\in[\chi]}$ be the set of canonical basis vectors so that $\{U e_{i,j}\}_{j\in[\chi]}$ covers all the $\{C_{i,j}\}_{j\in[\chi]}$.
Note that we may be able to cover them with less than $\chi$ canonical basis vectors, in that case we choose another canonical basis vector uniformly at random.
We shall show that the set $\cup_{i\in\cB''}\{e_{i,j}\}_{j\in[\chi]}$ is of size $|\cB''|\chi$ with probability $1-o(1)$, i.e. $\{e_{i,j}\}_{j\in[\chi]}$'s are mutually disjoint.

Fix any selection of $U$ over all columns of $\Pi$, which is a multiset of $d/\beta+d$ columns of $\Pi$. 
Note that this fixes all the $\{C_{i,j}\}_{i\in\cB'',j\in[\chi]}$ as well.
Then the relationship between $e_j$'s and the aforesaid multiset behaves in the following manner:
each $e_j$ chooses $1/\beta+1$ columns from the multiset uniformly at random, and any two distinct $e_j,e_k$ never choose the same column from the multiset.
Notice that 
\begin{enumerate*}[label=(\roman{*})]
    \item $|\{C_{i,j}\}_{i\in\cB'',j\in[\chi]}|=O(\chi 2^\ell\log^2s)=O(s^2\eps\log^2s)=O(\log^3d/\eps^2)$,
    \item for each $e_j$, $U e_j$ covers precisely $1/\beta+1=O(\log d/\eps)$ columns,
    \item $U$ chooses $d/\beta+d=\omega(\log^8d/\eps^6)$ columns in total,
\end{enumerate*}
any $e_j$ is associated with at most one column in $\{C_{i,j}\}_{i\in\cB'',j\in[\chi]}$ with probability $1-o(1)$ by a simple calculation.

Therefore, we conclude that with constant probability, there is a set $\cB''\subseteq[2m/s]$ of indices of submatrices such that 
\begin{enumerate}[label=(\arabic{*})]
    \item $|\cB''|=\Theta(2^\ell\log^2s)$; %
    \item $\cB''$ distributed over $[2m/s]$ uniformly at random without replacement;
    \item for each $i\in\cB''$, there is a set $\{e_{i,j}\}_{j\in[\chi]}$ of $\chi$ canonical basis vectors such that $\{U e_{i,j}\}_{j\in[\chi]}$ covers $\chi$ vectors in $\Pi_i$, and the $\chi$ columns are distributed uniformly at random over all the columns of $\Pi_i$ with replacement, the remaining $\chi/\beta$ columns are distributed uniformly at random over all the columns of $\Pi$ with replacement;
    \item $\cup_{i\in\cB''}\{e_{i,j}\}_{j\in[\chi]}$ is of size $|\cB''|\cdot\chi$, i.e. $\{e_{i,j}\}_{j\in[\chi]}$'s are mutually disjoint;
    \item $\{U e_{i,j}\}_{j\in[\chi]}$'s are mutually independent for different $i\in\cB''$.
\end{enumerate}

For simplicity, let $p_i \triangleq \Pr[\|X+\bar Y^{\ell}\|_2\notin (1\pm\eps)(\chi/\beta+\chi)|\cA_i]$.
Now we examine each of $\Pi U (e_{i,1}+\dots+e_{i,\chi})/\sqrt\chi$ for $i\in\cB''$.
Recall that none of the canonical basis vectors is examined more than once, therefore, $X$ and $Y$ from different $i\in \cB''$ are mutually independent by properties (4) and (5).
By properties (2) and (3), there exists $i\in\cB''$ such that $\|X+\bar Y^{\ell}\|_2 \notin (1\pm\eps)(\chi/\beta+\chi)$ with probability
\[
    1-\prod_{i\in \cB''}\left(1-p_i\right) \geq 1 - \prod_{i\in \cB''} e^{-p_i} = 1 - \exp\left(\sum_{i\in \cB''} p_i\right) =\Omega\left( \min\left\{1,\sum_{i\in \cB''} p_i\right\}\right).
\]
Define for each $i$ a random variable $X_i$ such that $X_i = p_i$ for $i\in \cB''$ and $X_i = 0$ otherwise, then 
$\sum_i X_i=\sum_{i\in \cB''}p_i$. 
The assumption \eqref{eqn:sum of p_i} states that $\E_{i\sim \Unif(\mathcal{B})} p_i = \Omega(1/(2^\ell \log^2 s))$.
Note that $X_i$'s are negatively correlated, and $\E[X_i]=\Omega({|\cB''|}p_i/{|\cB|})$, so $\E[\sum_{i\in \cB''}p_i]=\Omega(|\cB''|/2^\ell\log^2s)$.
As $|\cB''|=\Theta({2^{\ell}\log^2 s})$, by property (2) and \cref{claim: sum of negatively correlated}, %
$\sum_{i\in \cB''} p_i = \Omega(|\mathcal{B}'|/(2^\ell \log^2 s))$ with probability $1-o(1)$. %
Therefore, as long as there exists such a $\cB''$ of size $\Theta(2^\ell\log^2s)$, there is an $i\in \cB''$ such that $\|X+\bar Y^{\ell}\|_2\notin (1\pm\eps)(\chi/\beta+\chi)$ with constant probability.
Hence it must hold that
\begin{align*}
    |\cB'|&=O(2^\ell\log^2 s)
    \\
    |\cB|\cdot\binom{d/\beta+d}{\chi}\left(\frac{s}{m}\right)^\chi&=O({2^{\ell}\log^2 s})
    \\
    \frac{|\cB|s}{m}\left(\frac{d 2^{\ell}}{\chi}\right)^\chi\left(\frac{s}{m}\right)^{\chi -1}&=O({2^{\ell}\log^2 s})
    \\
    \frac{2^{\ell(\chi-1)}}{\log^2 s}\left(\frac{d}{\eps s}\right)^\chi s^{\chi -1}&=O(m^{\chi -1})
    \\
    m&=\Omega\left(\frac{2^{\ell}}{(s \log^2 s)^{1/(\chi-1)}}\cdot\left(\frac{d}{\eps}\right)^{\chi/(\chi-1)}\right). \qedhere
\end{align*}
    
\end{proof}

\subsubsection{The anticoncentration}
Conditioned on event $\cA$, we have $\|X\|_2^2\approx \chi+2^{\ell_1}\chi^2/s$.
We now examine $\langle X,\bar Y^{\ell}\rangle$ to find a good parameter $\beta$ to meet the conditions of Lemma~\ref{lem: distortion implies lower bound}.
Recall that $\chi=\lfloor 4\eps s\rfloor+4$, so $(1-1/\log s)\chi/s>2\eps+\eps^2+1/s$.

\paragraph{Case 1: $\ell_1\ge 3$ and $\Pr[\langle X,\bar Y^{\ell_1-4}\rangle\ge -(1/4)\cdot 2^{\ell_1}\chi^2/s|\cA]\ge 1/2$.}
By a union bound, 
\begin{enumerate*}[label=(\roman*)]
    \item $\langle X,\bar Y^{\ell_1-3}\rangle$ does not cancel $\|X\|_2^2 - (1+\eps)^2\chi$, and 
    \item $\|Y\|_2\ge (1-\eps)^2(\chi/\beta)$
\end{enumerate*}
happen simultaneously with probability $1/2-\delta$.
To summarize, if $U\sim \cD_\beta$ with $1/\beta=2^{\ell_1-3}$, conditioned on $\cA$, %
\begin{align*}
    \|X\|_2^2+2\langle X,Y\rangle+\|Y\|_2^2 &\ge (1-1/s)\chi+(2^{\ell_1}+1)\chi^2/s-2\cdot(1/4)\cdot 2^{\ell_1}\chi^2/s+(\chi/\beta)(1-\eps)^2\\
    &> (1-\eps)^2(\chi/\beta+\chi) + (1/2)\cdot 2^{\ell_1}\chi^2/s\\ 
    &\ge (1-\eps)^2(\chi/\beta+\chi) + 2(\chi/s)(\chi/\beta+\chi)\\ 	 %
    &> (1+\eps)^2(\chi/\beta+\chi)
\end{align*}
with probability at least $1/2-\delta$, where the last inequality is because $\chi/s>4\eps$.
It can be rephrased as 
\[
    \E_{i\sim\Unif(\cB)}\left[\Pr\left[ \left.\Norm{X+\bar Y^{\ell_1-3} }_2^2 > (1+\eps)^2(\chi/\beta+\chi) \right\vert \cA_i,\cA\right]\right]\ge \frac{1}{2}-\delta,
\]
which implies that
\[
    \E_{i\sim\Unif(\cB)}\left[\Pr\left[ \left.\Norm{X+\bar Y^{\ell_1-3}}_2^2 > (1+\eps)^2(\chi/\beta+\chi) \right\vert \cA_i\right]\right]=\Omega\left(\frac 1 {2^{\ell_1}\log^2 s}\right).
\]

\paragraph{Case 2: $\ell_1< 3$.}
We follow the argument in Case 1 but apply $U\sim\cD_\beta$ with $1/\beta=0$ to $\Pi$, then obtain that conditioned on $\cA$,
\[
    \|X\|_2^2\ge (1-1/s)\chi+\chi^2(1/s-1/s\log s)> (1+\eps)^2\chi,
\]
where the second inequality is because $(1-1/\log s)\chi/s>2\eps+\eps^2+1/s$. %
Then we conclude
\[
    \E_{i\sim\Unif(\cB)}\left[\Pr\left[ \left.\Norm{X}_2^2 > (1+\eps)^2\chi \right\vert \cA_i\right]\right]=\Omega\left(\frac 1 {2^{\ell_1}\log^2 s}\right).
\]

\paragraph{Case 3: $\ell_1\ge 3$ and $\Pr[\langle X,\bar Y^{\ell_1-3}\rangle < -(1/4)\cdot 2^{\ell_1}\chi^2/s|\cA]\ge 1/2$.}
It then follows that
\begin{align*}
 \Pr[\langle X,\bar Y^{\ell_1}\rangle< -2\cdot 2^{\ell_1}\chi^2/s] &= \E[ \Pr[\langle X,\bar Y^{\ell_1}\rangle< -2\cdot 2^{\ell_1}\chi^2/s]  \vert X] \\
 &\geq \E[ \Pr[\langle X,\bar Y^{\ell_1 - 3}\rangle< -(1/4)\cdot 2^{\ell_1}\chi^2/s]^{8}  \vert X] \\
 &\geq \E[ \Pr[\langle X,\bar Y^{\ell_1 - 3}\rangle< -(1/4)\cdot 2^{\ell_1}\chi^2/s]  \vert X]^{8} \\
 &= \Pr[\langle X,\bar Y^{\ell_1 - 3}\rangle< -(1/4)\cdot 2^{\ell_1}\chi^2/s]^{8} \\
 &\geq 1/256,
\end{align*}
where the second inequality follows from Jensen's inequality and the convexity of $x\mapsto x^8$.
By a union bound, with probability at least $1/256-\delta$, 
\begin{enumerate*}[label=(\roman*)]
    \item $\langle X,\bar Y^{\ell_1}\rangle$ dominates $\|X\|_2-(1-\eps)\chi$, and 
    \item $\|Y\|_2\le (1+\eps)(\chi/\beta-\chi)$
\end{enumerate*}
happen simultaneously.
To summarize, if $U\sim \cD_\beta$ with $1/\beta=2^{\ell_1}$, conditioned on $\cA$,
\begin{align*}
    \|X\|_2^2+2\langle X,Y\rangle +\|Y\|_2^2 &\le (1-1/s)\chi+(2^{\ell_1+1}+1)\chi^2/s-2\cdot 2\cdot 2^{\ell_1}\chi^2/s+(1+\eps)^2(\chi/\beta)
    \\
    & < (1+\eps)^2(\chi/\beta+\chi) - (2\cdot 2^{\ell_1}-1)\chi^2/s \\
    & \le (1+2\eps+\eps^2 - (5/3)\chi/s)(\chi/\beta+\chi) \\
    & < (1-3\eps+\eps^2)(\chi/\beta+\chi)
\end{align*}
with probability at least $1/256-\delta$, where the last inequality is because $\chi/s>4\eps$.
It can be rephrased as 
\[
    \E_{i\sim\Unif(\cB)}\left[\Pr\left[ \left.\Norm{X+\bar Y^{\ell_1} }_2^2 < (1-\eps)^2(\chi/\beta) \right\vert \cA_i,\cA\right]\right]\ge \frac{1}{256}-\delta,
\]
which implies that
\[
    \E_{i\sim\Unif(\cB)}\left[\Pr\left[ \left.\Norm{X+\bar Y^{\ell_1}}_2^2 < (1-\eps)^2(\chi/\beta) \right\vert \cA_i\right]\right]=\Omega\left(\frac 1 {2^{\ell_1}\log^2 s}\right).
\]

\subsection{Removing the assumption}\label{sec: removing assumption in denser case}
We find the most ``popular'' value of the entries of $\Pi$ with the following lemma.
\begin{lemma}[Refined from \cref{lem:theta}]
    Suppose that $\Pi$ is a matrix with column sparsity at most $s\ge 3$. %
    Then there exists a $\theta\ge (1-\eps)^2/s-1/s\log s$ such that %
    \begin{equation}%
        \E_j\left[ \sum_{k: |\Pi_{k,j}|^2\in[\theta,2\theta)} \Abs{\Pi_{k,j}}^2 \;\middle\vert\; \Norm{\Pi_{\star,j}}_2= 1\pm\eps\right]=\Omega(1/\log^2 s). 
    \end{equation}
\end{lemma}
We apply the preceding lemma to $\Pi$, then let $\chi\triangleq\lfloor (4\eps+\theta)/\theta(1-1/\log s)\rfloor+1\ge 2$.

Let distribution $\widetilde{\cD}$ be a mixture of $\cD_\beta$ and $\cD'_\beta$, which are parameterized by $\beta$, on $n\times d$ matrices.
With probability $1/3$, $\widetilde{\cD}=\cD_1$; with probability $1/3$, $\widetilde{\cD}$ is a $\cD'_\beta$ for $\beta \sim \Unif([\lceil\log s\rceil])$; and with probability $1/3$, $\widetilde{\cD}$ is a $\cD_\beta$ for $\beta\sim \Unif([\lceil\log s\rceil])$.

We shall prove the following theorem in this subsection.
\begin{theorem}\label{thm: main2 formal}
    Suppose that
    \begin{enumerate*}[label=(\roman*)]
    \item $s=O(\eps^{-1}\log d)$ and $s\geq 3$, and
    \item 
    $\Pi$ is an OSE for $U\sim\widetilde{\cD}$.
    \end{enumerate*}
    Then $\Pi$ must have at least
    \[
        m=\Omega\left(\frac{\theta^{1/(\chi-1)}}{s^{8\gamma/(\chi-1)}\log^{6/{(\chi-1)}+2}s}\cdot\left(\frac{d}{\eps}\right)^{\chi/(\chi-1)}\right)
    \]
    rows, where $\chi = \lfloor (4\eps+\theta)/(\theta(1-1/\log s))\rfloor+1\ge 2$ and $\theta\ge (1-\eps)^2/s-1/s\log s$.
\end{theorem}
Note that $4\eps/\theta = 4\eps s (1+2\eps + O(\eps^2) + O(1/\log s))$.
Thus for $\eps\le 1/202$ and $s=\Omega(\log(1/\eps)/\eps)\cap O((\log d)/\eps)$, the lower bound is at least
\[
m=\Omega\left(\left(\frac{d}{\eps}\right)^{1+\frac{1}{4\eps(1+\eps)s}}\cdot\frac{1}{s^{O(\delta)}}\right).
\]

\subsubsection{Partitioning OSE matrix}
Since $\Pi$ is an OSE for $U\sim\widetilde{\cD}$, $\Pi$ is also an OSE for $U\sim\cD_1$.
Thus $\|\Pi_{\star,j}\|_2=1\pm\eps$ for at least $n(1-3\delta/d)$ columns $j$ of $\Pi$ by \cref{claim: nonzeroes2}. 
\begin{lemma}[Refined from {\cite[Lemma 6]{LL22}}]\label{claim: nonzeroes2}
If $\Pi\in\mathbb{R}^{m\times n}$ is an $(\epsilon,\delta)$-subspace-embedding for $U\sim\widetilde\cD$, then $(1-3\delta/d)$-fraction of the columns of $\Pi$ has $\ell_2$-norm of $1\pm\epsilon$.
\end{lemma}

Now we partition $\Pi$ with new parameters.
\begin{claim}[Refined from \cref{claim: partition bucket}]\label{claim: partition bucket 2}
    Assume $\Pi\in\mathbb R^{m\times n}$ is of column sparsity $s\ge 3$. 
    $\Pi$ can be partitioned into $0.05 m \theta \log^2 s$ submatrices $\Pi_1,\dots,\Pi_{0.05 m\theta\log^2 s}\in\mathbb R^{m\times 20n/(m\theta\log^2 s)}$ such that for each $\Pi_i$ with $i\le 0.024m\theta$
   \begin{enumerate*}[label=(\arabic{*})]
    \item every column has $\ell_2$-norm $1\pm\eps$, and
    \item
    there is a row $r$ such that every entry in row $r$ 
    \begin{enumerate*}[label=(\roman{*})]
        \item 
    is in $[\sqrt\theta,\sqrt{2\theta}]$ and
    \item has the same sign.
    \end{enumerate*}
   \end{enumerate*} 
\end{claim}

For each $i\le 0.024m\theta$, let $\Pi'_i$ denote the set of columns obtained from the set of columns in $\Pi_i$ after removing row $r$. %
We apply \cref{lem:k good inner product} to each $\Pi'_i$ for $i\le 0.024m\theta$ with $k=\chi$, $\kappa=\theta/\log s$, $\Delta=\theta$, and obtain an $\ell$ for each $\Pi_i'$. Let $\ell_1$ be the majority of these $0.024m\theta$ values of $\ell$ and $\cB\subseteq[0.024m\theta]$ the index set of $i$'s such that \cref{lem:k good inner product} returns $\ell_1$ when applied to $\Pi'_i$ with $i\le 0.024m\theta$.
Note that $|\cB|=\Omega( m\theta/\log^2 s)$.
Let $X$ be a random variable which is obtained by sampling $i\in\cB$ uniformly at random then summing over $\chi$ uniformly random column vectors in $\Pi_i$, $X'_1,\dots,X'_\chi$ the $\chi$ random column vectors with zeroing out $r$-th row, $x_1,\dots,x_\chi$ the values of the $r$-th row in $\Pi_i$.
Then 
\[
    \|X\|_2^2=\sum_{i,j\in [\chi]}(\langle X'_i,X'_j\rangle+x_ix_j)=\sum_{i,j\in[\chi]:i\ne j}(\langle X'_i,X'_j\rangle+x_ix_j)+\sum_i(\|X'_i\|_2^2+x_i^2).
\]
Recall that each column in $\Pi_i$ is of $\ell_2$-norm $1\pm \eps$, so $\sum_i(\|X'_i\|_2^2+x_i^2)=(1\pm \eps)^2\chi$.
Recall that each entry in row $r$ and in $\Pi_i$ is in $[\sqrt\theta,\sqrt{2\theta}]$ and has the same sign, so $\sum_{i\ne j}x_ix_j\in[\chi(\chi-1)\theta,2\chi(\chi-1)\theta]$.
Finally, $\sum_{i,j\in[\chi]:i\ne j}(\langle X'_i,X'_j\rangle$ is characterized by \cref{lem:k good inner product}.
Let $\cA$ be the event that \[(1-\eps)^2\chi-\theta\chi+\chi^2(\theta-\theta/\log s)\le \|X\|_2^2\le (1+\eps)^2\chi-\theta\chi+3\chi^2\theta\] if $\ell_1=-1$, or \[(1-\eps)^2\chi-\theta\chi+(2^{\ell_1}+1)\chi^2\theta < \|X\|_2^2\leq (1+\eps)^2\chi-\theta\chi+(2^{\ell_1+1}+2) \chi^2\theta\] if $\ell_1\ge 0$.
Note that $\Pr[\cA]=\Omega(1/(2^{\ell_1}\log^2 s))$ if $i\in \cB$ by \cref{lem:k good inner product} and \cref{claim: partition bucket 2}.

\subsubsection{Obtaining lower bound via anticoncentration}
Let $\cA_i$ be the event that $U$ chooses at least $\chi$ columns in $\Pi_i$, $Y'$ a uniformly random column over $\Pi$, $\bar Y^\ell$ the sum of $\chi 2^\ell$ independent copies of $Y'$.
We are going to invoke the following lemma to complete the proof.
\begin{lemma}\label{lem: general distortion implies lower bound}
    Assume that $s=O(\eps^{-1}\log d)$, $s\ge 3$, and $d\ge 1/\eps^7$.
If there exists $\ell,1/\beta\ge 0$ such that $\Pi$ is an OSE for $\cD'_\beta$ and
\begin{equation}
\E_{i\sim\Unif(\cB)}\left[ \Pr\left[ \left.\Norm{X+\bar Y^{\ell}}_2^2 \notin (1\pm\eps)^2(\chi/\beta+\chi) \right\vert \cA_i \right] \right] = \Omega\left(\frac 1 {2^{\ell}\log^2 s}\right),
\end{equation}
then it must hold that 
\[
    m=\Omega\left(\frac{2^{-\ell/(\chi-1)}(1+1/\beta)^{\chi/(\chi-1)}}{\theta^{-1/(\chi -1)}\log^{6/{(\chi-1)}+2} s}\cdot\left(\frac{d}{\eps}\right)^{\chi/(\chi-1)}\right).
\]
\end{lemma}
\begin{proof}
    The proof is almost identical with the proof of \cref{lem: distortion implies lower bound}, except the calculations. In this case, we have
\[
    |\cB'|=\Theta\left( |\cB|\cdot\binom{d(1+1/\beta)}{\chi}\left(\frac{1}{m\theta\log^2 s}\right)^\chi\right).
\]
Also, $\beta$ is not fixed to $2^{\ell_1}$, so
\begin{align*}
    |\cB|\cdot\binom{d(1+1/\beta)}{\chi}\left(\frac{1}{m\theta\log^2 s}\right)^\chi&=O({2^{\ell}\log^2 s})
    \\
    \frac{|\cB|}{m\theta\log^2 s}\left(\frac{d(1+1/\beta)}{\chi}\right)^\chi\left(\frac{1}{m\theta\log^2 s}\right)^{\chi -1}&=O(2^{\ell}\log^2 s)
    \\
    \frac{2^{-\ell}(1+1/\beta)^{\chi}}{\log^6 s}\left(\frac{d}{\eps/\theta}\right)^\chi \left(\frac{1}{\theta\log^2 s}\right)^{\chi -1}&=O(m^{\chi -1})
    \\
    m&=\Omega\left(\frac{2^{-\ell/(\chi-1)}(1+1/\beta)^{\chi/(\chi-1)}}{\theta^{-1/(\chi -1)}\log^{6/{(\chi-1)}+2} s}\cdot\left(\frac{d}{\eps}\right)^{\chi/(\chi-1)}\right). \qedhere
\end{align*}
\end{proof}

\subsubsection{The anticoncentration}
Recall that $\Pi$ is an $(\eps,\delta)$-subspace-embedding for $U\sim\widetilde{\cD}$.
Recall that we assume $n\ge K(sd)^2/\delta$ for large enough constant $K$ so that $U\sim\widetilde{\cD}$ is an isometry with probability $1-\delta/(2K)$.
By Markov's inequality, for $(1-\gamma)$-fraction of $\ell\in[L]$, $\Pi$ is an $(\eps,2\delta/\gamma)$-subspace-embedding for $U\sim\cD_{2^{-\ell}}$, where $\gamma\triangleq K \delta = O(\delta)$, so that $2\delta/\gamma=2/K$ is a small constant.
Similarly, for $(1-\gamma)$-fraction of $\ell\in[L]$, $\Pi$ is an $(\eps,2\delta/\gamma)$-subspace-embedding for $U\sim\cD'_{2^{-\ell}}$, where $\gamma\triangleq K \delta =O(\delta)$, so that $2\delta/\gamma=2/K$ is a small constant.

Conditioned on event $\cA$, we have $\|X\|_2\approx \chi+2^{\ell_1}\chi^2\theta$.
We now examine $\langle X,\bar Y^{\ell}\rangle$ to find a good parameter $\beta$ to meet the conditions of Lemma~\ref{lem: general distortion implies lower bound}.
Recall that $\chi=\lfloor (4\eps+\theta)/(\theta(1-1/\log s))\rfloor+1>4\eps/\theta$, so 
\begin{align}
    -\theta+\chi\theta-\chi\theta/\log s>4\eps.\label{eq: chi lower bound}
\end{align}

We follow the same argument as in the previous subsection.
To this end, we shall find an integer $\ell_2\le {\ell_1}$ such that $\Pi$ is an $(\eps,2/K)$-subspace-embedding for $U\sim\cD_{2^{\ell_2}}$, $U\sim\cD'_{2^{\ell_2}}$, $U\sim\cD_{2^{\ell_2-3}}$ and $U\sim\cD'_{2^{\ell_2-3}}$ simultaneously.
By the pigeonhole principle, for $\ell_1\ge 8\gamma\log s+3$, there is an integer $\ell_2\in[\ell_1-8\gamma\log s,{\ell_1}]$ which meets the preceding requirement.
Note that
\begin{align}
    2^{\ell_2}=\Omega( 2^{\ell_1}/s^{8\gamma}).\label{eq: ell' and ell 2}
\end{align}

\paragraph{Case 1: $\ell_1> 8\gamma\log s+3$ and $\Pr[\langle X,\bar Y^{\ell_2-3}\rangle\ge -(1/4)\cdot 2^{\ell_1}\chi^2\theta]\ge 1/2$.}
By our assumption, if we examine $\Pi U$ with $1/\beta=2^{\ell_2-3}\le 2^{\ell_1}/8$
\begin{align*}
    \|X\|_2^2+\|Y\|_2^2+2\langle X,Y\rangle \ge& (1-\eps)^2(\chi/\beta+\chi) -\theta\chi + (2^{\ell_1-1}+1)\chi^2\theta \\
    \ge&(1-2\eps+\eps^2+2(\chi\theta))(\chi/\beta+\chi)>(1+\eps)^2(\chi/\beta+\chi)
\end{align*}
with probability at least $1/2-2/K$, where the last inequality is because $\chi\theta>4\eps$.
So we have for $1/\beta=2^{\ell_2-3}\ge 2^{\ell_1}/8s^{8\gamma}$ that
\[
    \E_{i\sim\Unif(\cB)}\left[\Pr\left[ \left.\Norm{X+\bar Y^{\ell_1-3}}_2^2 > (1+\eps)^2(\chi/\beta+\chi) \right\vert \cA_i\right]\right]=\Omega\left(\frac 1 {2^{\ell_1}\log^2 s}\right).
\]
\paragraph{Case 2: $\ell_1\le 8\gamma\log s+3$.}
We follow the argument in Case 1 but apply $U\sim\cD'_\beta$ with $1/\beta=0$ to $\Pi$, then obtain that conditioned on $\cA$,
\[
    \|X\|_2^2\ge (1-\eps)^2\chi-\theta\chi+\chi^2(\theta-\theta/\log s)> (1+\eps)^2\chi,
\]
where the second inequality is due to \cref{eq: chi lower bound}.
Then we conclude for $1/\beta=0\ge 2^{\ell_1}/8s^{8\gamma}-1$
\[
    \E_{i\sim\Unif(\cB)}\left[\Pr\left[ \left.\Norm{X}_2^2 > (1+\eps)^2\chi \right\vert \cA_i\right]\right]=\Omega\left(\frac 1 {2^{\ell_1}\log^2 s}\right).
\]
\paragraph{Case 3: $\ell_1> 8\gamma\log s+3$ and $\Pr[\langle X,\bar Y^{\ell_2-3}\rangle< -(1/4)\cdot 2^{\ell_1}\chi^2\theta]\ge 1/2$.}
By our assumption, if we examine $\Pi U$ with $1/\beta=2^{\ell_2}$
\begin{align*}
    \|X\|_2^2+\|Y\|_2^2+2\langle X,Y\rangle \le& (1+\eps)^2(\chi/\beta+\chi) -\theta\chi - (2\cdot 2^{\ell_1}-2) \chi^2\theta\\
    \le&(1+2\eps+\eps^2-(5/3)(\chi\theta))(\chi/\beta+\chi)<(1-\eps)^2(\chi/\beta+\chi)
\end{align*}
with probability at least $1/256-2/K$, where the last inequality is because $\chi\theta>4\eps$.
So we have for $1/\beta=2^{\ell_2}\ge 2^{\ell_1}/s^{8\gamma}$ that
\[
    \E_{i\sim\Unif(\cB)}\left[\Pr\left[ \left.\Norm{X+\bar Y^{\ell_1}}_2^2 < (1-\eps)^2(\chi/\beta+\chi) \right\vert \cA_i\right]\right]=\Omega\left(\frac 1 {2^{\ell_1}\log^2 s}\right).
\]

Applying \cref{lem: general distortion implies lower bound} with the values of $\beta$ in Cases 1 and 3, we obtain
\[
    m=\Omega\left(\frac{2^{-\ell_1/(\chi-1)}(1+2^{\ell_2})^{\chi/(\chi-1)}}{\theta^{-1/(\chi -1)}\log^{6/{(\chi-1)}+2} s}\cdot\left(\frac{d}{\eps}\right)^{\chi/(\chi-1)}\right)
\]
which by \cref{eq: ell' and ell 2} is at least
\[
    m=\Omega\left(\frac{2^{\ell_1}\cdot\theta^{1/(\chi-1)}}{s^{8\gamma\chi/(\chi-1)}\log^{6/{(\chi-1)}+2}s}\cdot\left(\frac{d}{\eps}\right)^{\chi/(\chi-1)}\right).
\]

In Case 2, we have
\[
    m=\Omega\left(\frac{\theta^{1/(\chi -1)}}{(8s^{8\gamma})^{1/(\chi-1)}\log^{6/{(\chi-1)}+2} s}\cdot\left(\frac{d}{\eps}\right)^{\chi/(\chi-1)}\right).
\]

\bibliography{refs}
\bibliographystyle{alpha}

\appendix
\section{Incompatibility with the previous approaches} \label{sec:incompatibility}

We now explain why \cref{eqn: intuition} is incompatible with the existing strategy.
Recall the strategy used by \cite{NN14,LL22}, described in \cref{sec: previous}. All existing arguments prove that the following two quantities are large simultaneously:
\begin{enumerate}[label=(\roman*)]
	\item the number of column pairs $(\Pi V)_{\star,i},(\Pi V)_{\star,j}$ such that the two columns are approximately chosen uniformly at random from some set of columns of $\Pi$  with guarantee that  
	there is a row $k$ such that $(\Pi V)_{k,i},(\Pi V)_{k,j}\ge \sqrt \eps$; \footnote{Or $\le -\sqrt\eps$, we assume $\ge \sqrt\eps$ without loss of generality.}
	\item the probability that each column pair in (i) has a large inner product.
\end{enumerate}
Previous works have the guarantee for (ii), namely, \cref{eqn: previous intuition}. Hence the problem reduces to showing that the quantity (i) is large. It is tempting to obtain a higher lower bound  by replacing the  guarantee for (ii) in the previous works  with \cref{eqn: intuition}; however, this will not work because \cref{eqn: intuition} is not compatible with the previous proofs for lower bounding (i). 

The lower bound for quantity (i) in \cite{NN14} is too small for our purpose, so we examine the argument in \cite{LL22}, which proceeds in the following greedy manner:
\begin{enumerate}[label=(\arabic*)]
	\item Maintain an index set $S$ of the ``remaining'' columns of $\Pi$, initialized to $S=[n]$, together with an index set $S'$ of the   the ``remaining'' column vectors of $\Pi V$, initialized to $S'=[d/\beta]$, where $d/\beta$ is the number of columns of $V$.
	\item Maintain the following invariant: for all $v\in S'$, the column $(\Pi V)_{\star,v}$ is uniformly distributed over $\{\Pi_{\star,i}\}_{i\in S}$.
	\item Pick an arbitrary column $v$ from $S'$. Check if there is another vector $u$ in $S'$ such that $(\Pi V)_{k,u},(\Pi V)_{k,v}\ge \sqrt \eps$ for some row $k$.%
	\item If such $k$ exists,  choose such a $v$ uniformly at random, include $(u,v)$ in the column pairs for quantity (i), and remove $u,v$ from $S'$;
	\item Otherwise, all the other vectors $v$ in $S'$ do not have a row $k$ such that $(\Pi V)_{k,u},(\Pi V)_{k,v}\ge \sqrt \eps$.
	Remove $u$ from $S'$ and remove some columns from $S$ so that the invariant (2) holds.
\end{enumerate}
To obtain the lower bound for (ii), {\cite[Lemma 16]{LL22}} shows that the distribution of the column pairs obtained in Step (4) can be approximated by a distribution which is easier to analyze, from which we sample a row $k$ and then sample two columns from $\left\{i\in S:\Pi_{k,i}\ge\sqrt\eps\right\}$. 
{\cite[Lemma 9]{NN14}} and {\cite[Lemma 3]{LL22}} show that \cref{eqn: previous intuition} holds for \emph{any} row $k$ and \emph{any} column set $S$.

To see the compatibility issue of replacing (ii) with \cref{eqn: intuition}, note that $S$ keeps \emph{changing} over the whole greedy procedure because of Step (5).
By the invariant in Step (2), the joint distribution of the columns $u,v$ obtained in Step (4) is \emph{changing} correspondingly.
If we apply \cref{eqn: intuition} with $X=\left\langle(\Pi'_k V)_{\star,u},(\Pi'_k V)_{\star,v}\right\rangle$, then the distribution of $X$, in other words, the $\ell$ in \cref{eqn: intuition}, will also be \emph{changing}.
But, we have to fix the parameter $d/\beta$ for analyzing $\Pi V$, which is chosen according to $\ell$, before going through the greedy procedure. 
One may expect to fix $d/\beta$ according to the most popular $\ell$ among the $\ell$s  determined in Step (5).
However, the set of the columns which are likely removed from $S$ in Step (5) depends on the value of $d/\beta$,\footnote{For two vectors $u,v$, we denote statement `` there is a coordinate $k$ such that $u_k,v_k\approx\sqrt\eps$ '' by
	$u\overset\eps\approx v$.
	Indeed, for an $x\in S$ with $\Pr_{Y}[\Pi_{\star,x}\overset{\eps}{\match} \Pi_{\star,Y}]=\omega(1/|S'|)$, there exists $y\in S'$ such that $\Pi_{\star,x}\overset{\eps}{\match} \Pi_{\star,y}$ with probability $1-o(1)$; for $x$ with $\Pr_{Y}[\Pi_{\star,x}\overset{\eps}{\match} \Pi_{\star,Y}]= O(1/|S'|)$, there exists such a $y$ with probability $\Theta(|S'|\cdot\Pr_{Y}[\Pi_{\star,x}\overset{\eps}{\match} \Pi_{\star,Y}])$. The distribution of the columns $v\in S'$ which fail the check in Step (3) depends on the aforementioned probabilities, i.e. the probabilities that there exists such a $y$ for each $x\in S$. On the one hand, the distribution of the columns removed from $S$ in Step (5) depends on the distribution of failed $v\in S'$ in Step (3). On the other hand, the aforementioned probabilities depend on $|S'|\approx d/\beta$.} and so does the value of the most popular $\ell$.

Now we have hit a dead end and made no progress. A radically different approach, as explained in \cref{sec:our_approach}, was therefore pursued.

\section{Proof of \cref{claim: sum of negatively correlated}} \label{sec:sum of negatively correlated}
Let $p_i\triangleq \Pr[X_i=a_i]$.
Since $X_i$'s are negatively correlated, and $a_i\le 1$ for all $i$,
\[
    \Var\left[\sum_iX_i\right]\le \sum_i\Var[X_i]=\sum_i (a_i^2p_i-a_i^2p_i^2)\le \sum_ia_ip_i=\E\left[\sum_iX_i\right].
\]
By Chebyshev's inequality,
\[
    \Pr\left[\left|\sum_i X_i-\mu\right|\ge \mu/2\right]\le 4\cdot\Var\left[\sum_i X_i\right]/\mu^2\le 4/\mu.
\]

\section{Proof of \cref{lem:k good inner product}} \label{sec:k good inner product}
Let $X\triangleq\sum_{i\ne j\in[k]}\langle v_i,v_j\rangle$ for simplicity.
Note that if $\Pr[X\ge -k^2\kappa]\ge 1/2$ there is nothing to prove, we henceforth assume that $\Pr[X< -k^2\kappa]\ge 1/2$.

As $\E[X]\ge 0$, by the law of total expectation
\[
    \E[X|X\ge 0]\cdot\Pr[X\ge 0]\ge -\E[X|X<0]\cdot\Pr[X<0]\ge k^2\kappa/2.
\]
Since $X\le (1+\eps)^2k^2$, we have that
\begin{align*}
    &\E[X|X\in[0,k^2\Delta]]\cdot\Pr[X\in[0,k^2\Delta]]\\
    &\quad+\sum_{i=0}^{\lceil\log((1+\eps)/\Delta)\rceil-1}\E[X|X\in(2^ik^2\Delta,2^{i+1}k^2\Delta]]\cdot\Pr[X\in(2^ik^2\Delta,2^{i+1}k^2\Delta]]\ge k^2\kappa/2.
\end{align*}
By the pigeonhole principle, at least one of the $\lceil\log((1+\eps)/\Delta)\rceil+1$ terms on the LHS is at least $k^2\kappa/2(\lceil\log((1+\eps)/\Delta)\rceil+1)$.
If it is the first term in the preceding inequality, we conclude with (i); otherwise, we conclude with (ii).

\end{document}